\documentclass[sigconf,authorversion]{acmart}

\copyrightyear{2026}
\acmYear{2026}
\setcopyright{cc}
\setcctype{by}
\acmConference[CHI '26]{Proceedings of the 2026 CHI Conference on Human Factors in Computing Systems}{April 13--17, 2026}{Barcelona, Spain}
\acmBooktitle{Proceedings of the 2026 CHI Conference on Human Factors in Computing Systems (CHI '26), April 13--17, 2026, Barcelona, Spain}
\acmDOI{10.1145/3772318.3791062}
\acmISBN{979-8-4007-2278-3/2026/04}

\usepackage{xcolor}
\usepackage{enumitem} 
\usepackage{graphicx}
\usepackage{booktabs}
\usepackage{array,makecell}

\usepackage{subcaption}
\usepackage{framed}

\usepackage{soul}
\usepackage[inkscapelatex=false]{svg}

\definecolor{ao(english)}{rgb}{0.0, 0, 1}

\begin{document}

\title{Learning from AVA: Early Lessons from a Curated and Trustworthy Generative AI for Policy and Development Research}

\author{Nimisha Karnatak}
\authornote{Work done during internship at The World Bank (nkarnatak@worldbank.org)}
\affiliation{%
  \institution{University of Oxford}
  \city{Oxford}
  \country{United Kingdom}
}
\email{nimisha.karnatak@some.ox.ac.uk}

\author{Mohamad Chatila}
\affiliation{%
  \institution{The World Bank Group}
  \city{Washington}
    \state{DC}
  \country{USA}}
\email{mchatila@worldbank.org}

\author{Daniel Alejandro Pinzón Hernández}
\affiliation{%
  \institution{The World Bank Group}
  \city{Washington}
    \state{DC}
  \country{USA}}
\email{dpinzonhernandez@worldbank.org}

\author{Reza Yazdanfar}
\affiliation{%
  \institution{Nouswise, Inc.}
  \city{Dover}
  \state{DE}
  \country{USA}}
\email{reza@nouswise.com}

\author{Michelle Dugas}
\affiliation{%
  \institution{The World Bank Group}
  \city{Washington}
   \state{DC}
  \country{USA}}
\email{mdugas@worldbank.org}

\author{Renos Vakis}
\affiliation{%
  \institution{The World Bank Group}
  \city{Washington}
  \state{DC}
  \country{USA}}
\email{rvakis@worldbank.org}

\begin{abstract}
General-purpose LLMs pose misinformation risks for development and policy experts, lacking epistemic humility for verifiable outputs. We present AVA (AI + Verified Analysis), a GenAI platform built on a curated library of over 4,000 World Bank Reports with multilingual capabilities. AVA’s multi-agent pipeline enables users to query and receive evidence-based syntheses. It operationalizes epistemic humility through two mechanisms: citation verifiability (tracing claims to sources) and reasoned abstention (declining unsupported queries with justification and redirection). 
\noindent
We conducted an in-the-wild evaluation with over 2,200 individuals from heterogeneous organisations and roles in 116 countries, via log analysis, surveys, and 20 interviews. Difference-in-Differences estimates associate sustained engagement with 2.4–3.9 hours saved weekly. Qualitatively, participants used AVA as a specialized “evidence engine”; reasoned abstention clarified scope boundaries, and trust was calibrated through institutional provenance and page-anchored citations. 
We contribute design guidelines for specialized AI and articulate a vision for `ecosystem-aware' Humble AI.
\end{abstract}



\begin{teaserfigure}
    \centering
    \includegraphics[width=0.9\linewidth]{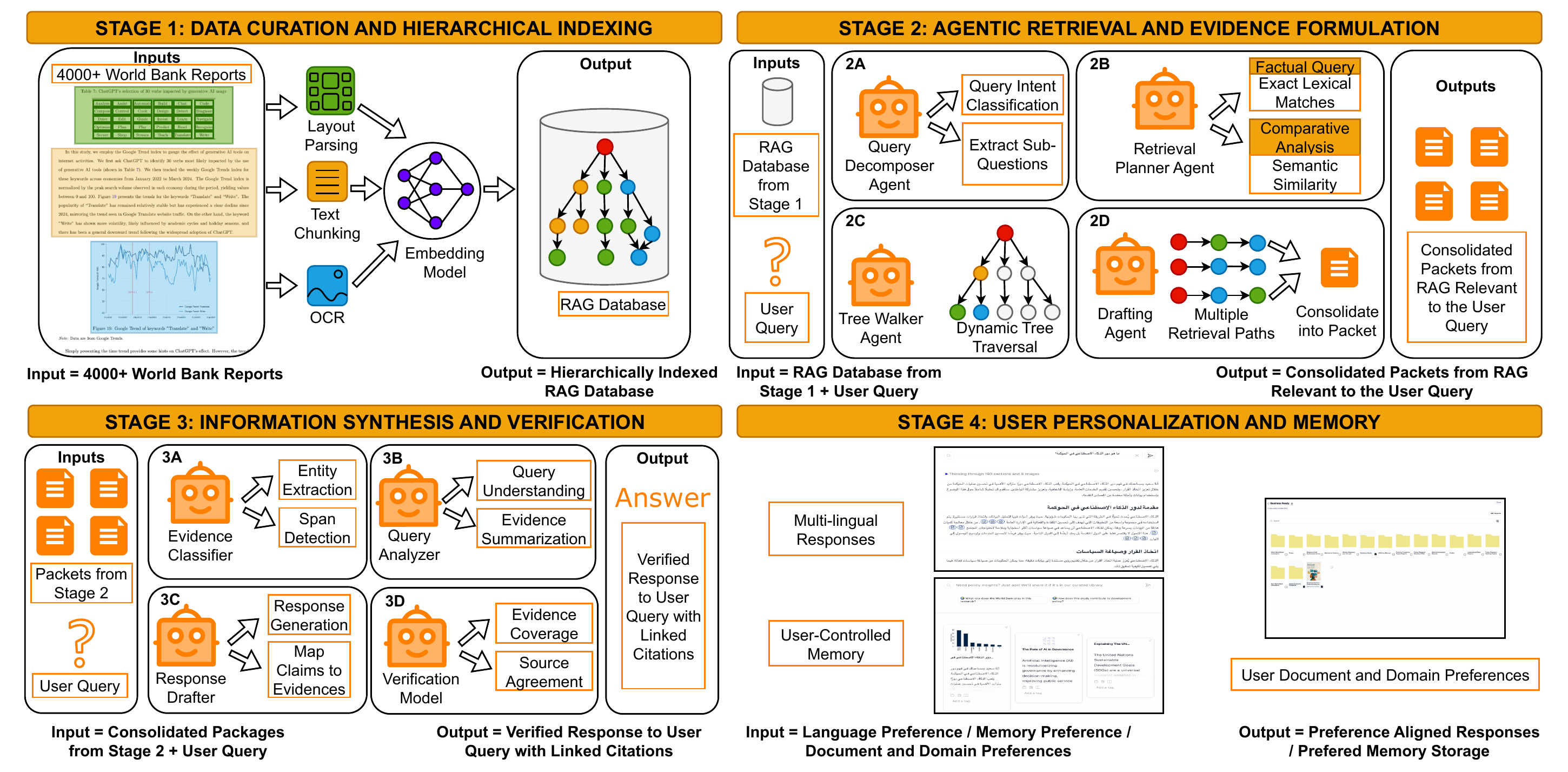}
    \caption{AVA System Architecture: Stage 1 curates 4000+ World Bank Reports into a hierarchical RAG; Stage 2 uses agentic retrieval (decomposition, planning, tree-walking, drafting); Stage 3 synthesizes and verifies evidence to produce citation-grounded answers or reasoned abstention; Stage 4 delivers multilingual, preference-aligned responses with interactive citations.}
    \Description{The image shows a four-stage architecture of the AVA system, arranged from left to right. Stage 1: Data curation and indexing. Over 4,000 World Bank reports are collected and processed. The system parses document layouts, extracts text using OCR, splits the text into chunks, and converts them into embeddings. These are organized into a hierarchically indexed retrieval-augmented generation (RAG) database. Stage 2: Agentic retrieval and evidence formulation. A user query enters the system and is handled by multiple specialized agents. One agent decomposes the query and identifies sub-questions. Another plans retrieval strategies, including exact keyword matching, semantic similarity, and comparative analysis. A tree-walking agent explores multiple retrieval paths through the database, and a drafting agent consolidates the results into evidence packets relevant to the query. Stage 3: Information synthesis and verification. The consolidated evidence packets are analyzed to extract entities, detect relevant text spans, and summarize evidence. The system maps claims to sources and verifies coverage and agreement across documents. The output is a verified answer with linked citations, or a reasoned abstention if evidence is insufficient. Stage 4: User personalization and memory. The final response is adapted to the user’s language, document preferences, and domain interests. The system supports multilingual outputs, interactive citations, and user-controlled memory for storing preferences or past interactions. Overall, the diagram illustrates how AVA moves from large-scale document ingestion to agent-driven retrieval, evidence verification, and personalized, citation-grounded responses.}
    \label{fig:AVA_architecture}
\end{teaserfigure}

\maketitle

\section{Introduction}

Public policy and development researchers navigate large, heterogeneous document corpora, often under significant time pressure. While Large Language Models (LLMs) can accelerate synthesis, their application in high-stakes domains such as policy-making is constrained by a lack of verifiable sources and a tendency to fabricate information, commonly referred to as hallucination~\citep{Leiser2024HILL,Cheng2024RELIC,Lee2024PaperWeaver,Sun2024MetaWriter}. Prior HCI research has established that user trust is contingent upon inspectable evidence and clearly communicated system boundaries \cite{wester2024denials,rapp2025chatgpt}. This has motivated interface designs that surface data provenance for verification and incorporate intentional refusal states when an answer cannot be substantiated \cite{venkit2024searchengines,rahman2025hallucinationtruthreviewfactchecking,liao2023aiTransparency}. These practical interface strategies represent operationalizations of epistemic humility, defined as a system’s capacity to recognise and communicate its own fallibility and limits~\cite{knowles2023humble,nair2025humble,tong2025measuring,ordonez2025humility}, and are commonly framed within the Humble AI paradigm, which emphasizes resisting over-claiming and prioritizing user verification over blind reliance~\cite{knowles2023humble,nair2025humble, celi2025teaching}.

While recent work has begun to translate Humble AI principles\footnote{Knowles et al.~\cite{knowles2023humble} define Humble AI via three principles: skepticism, curiosity  and commitment. See Sec 2.1 for details.}into concrete interface mechanisms, such as uncertainty-aware ranking in hiring~\cite{nair2025humble}, these initial efforts have predominantly focused on predictive tasks that classify or rank existing data, within controlled or short-horizon studies. This leaves two critical gaps. First, as generative AI models enter professional workflows, understanding how humility mechanisms function in systems that produce new text, rather than merely score or rank existing data as in predictive AI, poses distinct theoretical and design challenges (for example, hallucination vs misclassification)~\cite{nair2025humble}. Second, we lack longitudinal, in-the-wild evidence on how such systems integrate into everyday professional practice~\cite{pang2025llmification,wang2024task}.

\begin{figure*}
    \centering
    \includegraphics[width=0.7\linewidth]{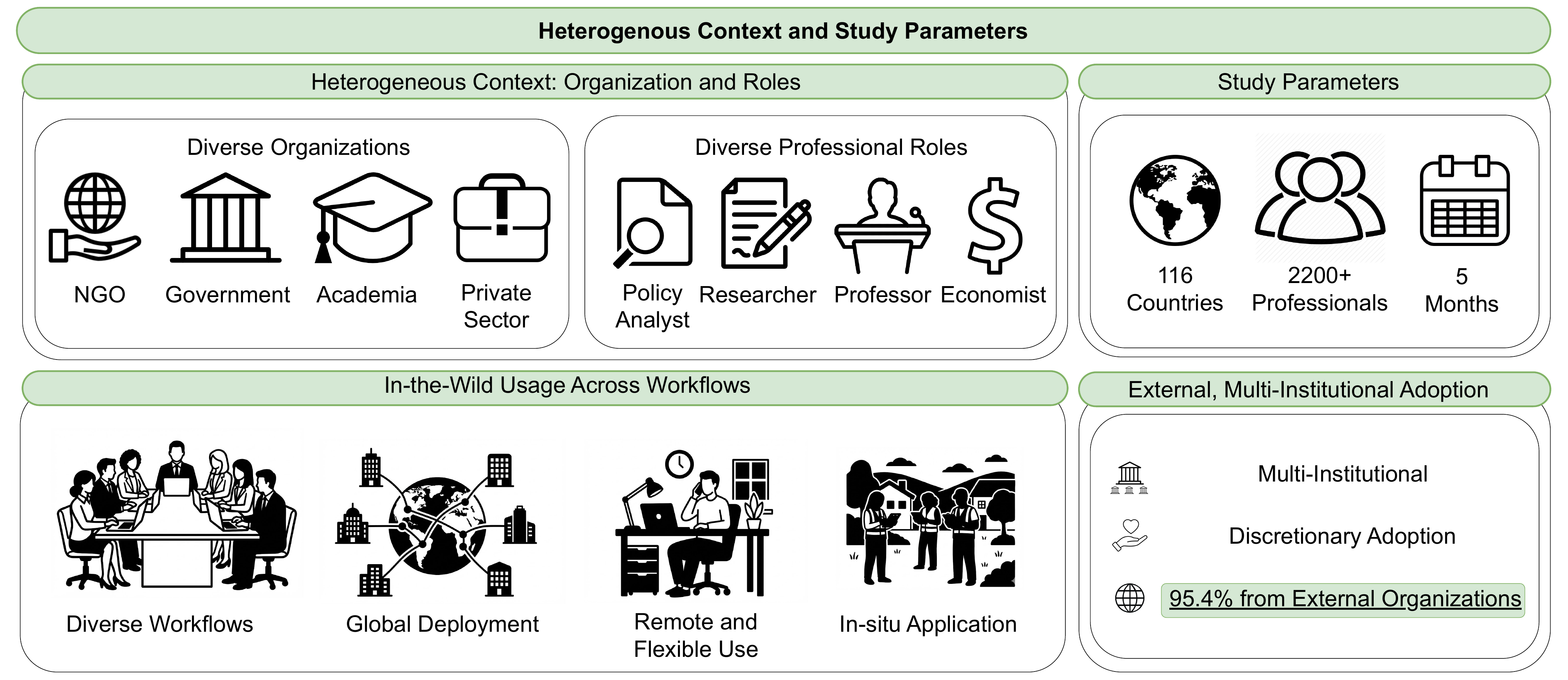}
    \vspace{-10pt}
     \caption{AVA's multi-institutional external deployment across 116 countries. Over 2,200 professionals used the system over five months, with 95.4\% employed by organizations external to the host institution (NGOs, academia, government professionals, etc).
    \vspace{-10pt}}
    \Description{The image summarizes AVA’s real-world deployment context and study parameters. At the top, the diagram highlights heterogeneous organizational and professional contexts. AVA was used by individuals from diverse organizations, including non-governmental organizations, government agencies, academic institutions, and the private sector. The professional roles represented include policy analysts, researchers, professors, and economists. To the right, the study scale is shown with three key metrics: deployment across 116 countries, usage by more than 2,200 professionals, and a study duration of five months. Along the bottom, the image illustrates in-the-wild usage across workflows, emphasizing global deployment, remote and flexible use, collaborative work settings, and in-situ application within everyday professional environments. The final section highlights external, multi-institutional adoption, noting that 95.4 percent of users came from organizations external to the host institution, indicating discretionary, voluntary uptake rather than mandated internal use. Overall, the image conveys that AVA was adopted globally, across institutions and roles, and used in real professional workflows at scale.}
    \Description{A brief description of the image for accessibility.}
    \label{fig:multi-institional}
\end{figure*}

To address these gaps, we present \textbf{\textsc{AVA} (AI + Verified Analysis)}, a source-linked, evidence-bounded, multi-agent generative AI assistant for policy and development professionals~\footnote{In this paper, “policy and development professionals” refers to practitioners working in international policy and development contexts (e.g., economic policy and public-sector reform) across NGOs, government agencies, academia, and international organizations. (See Fig.\ref{fig:multi-institional}, Table\ref{tab:participant-demographics}, Sec. 4.1 for details)}. For any query, AVA (i) returns synthesized answers with page-level, clickable citations and in-context highlighting; (ii) issues reasoned abstention, explicitly responding ``I don’t know'' with a brief rationale and reformulation paths when the corpus cannot substantiate a claim; and (iii) provides an in-place drafting and editing workspace. AVA is deployed over a curated library of 4,000+ official World Bank Reports and supports queries in over 60 languages (See appendix \ref{app:langauge}). In this paper, we situate AVA within CSCW and HCI accounts of how new technologies become embedded in professional practice~\cite{suchman1987plans,orlikowski2000using}. We use this lens to investigate whether operationalizing epistemic humility through source-linked answers and scope-awareness can enhance real-world evidence~\footnote{Appendix \ref{appendix:humility} details the operationalization strategies and design rationale.}.

Prior work characterizes integration of technology in workflow through appropriation \cite{dourish2003appropriation}, trust calibration \cite{lee2004trust}, sustained use \cite{orlikowski1995evolving}, as generative AI enters high-stakes workflows, debates about whether and how its use should be disclosed have gained prominence~\cite{hryciw2023guiding, hosseini2023ethics,eu_ai_act_2024}. Within this landscape, disclosure operates as a boundary practice through which professionals define the scope of their responsibility and negotiate professional and institutional accountability \cite{bahammam2025transparency}. Drawing on these four central dimensions from CSCW and HCI literature \cite{orlikowski2000using,yang2019unremarkable,xiao2025ai} we investigate AVA’s sociotechnical integration into professional workflows through questions that map to the gaps identified above:
\begin{itemize}
    \item \textbf{RQ1: Human–AI collaboration (Appropriation \& Use).} How do policy professionals appropriate AVA within evidence-based workflows, and what interaction patterns emerge across tasks? \textit{(Addressing Gap 2: lack of empirical evidence on AI integration into everyday professional practice.)}
    
    \item \textbf{RQ2: System boundaries (Refusal \& Reliance).} How do verification mechanisms, including citations and reasoned abstention, shape trust calibration and users’ mental models? \textit{(Addressing Gap 1: limited understanding of how Humble AI mechanisms function in generative systems as opposed to predictive ones.)}
    
    \item \textbf{RQ3: Sustained value (Retention \& Efficiency).} To what extent does sustained use predict perceived efficiency gains, and which first-session behaviors distinguish returning users? \textit{(Addressing Gap 2: lack of longitudinal evidence.)}
    
    \item \textbf{RQ4: Disclosure \& accountability (Authorship and \\Norms).} How do professionals reason about and enact disclosure of AVA’s assistance in their workflows? \textit{(Addressing policy-specific requirements for transparency and accountability.)}
\end{itemize}

\noindent
To answer these questions, we conducted a five-month, global, multi-institutional, in-the-wild deployment of AVA using a mixed-methods approach (see Figure~\ref{fig:multi-institional} for deployment scope and context).Crucially, the study engaged over 2,200 professionals across 116 countries. Of these, 95.4\% were employed by organizations external to the host institution, spanning NGOs, government professionals, academia, and the private sector. Quantitatively, we analyzed usage logs linked to baseline and endline surveys (2,259 registrants; 3,797 queries; matched $n=1,029$ across 116 countries). Qualitatively, we conducted 20 semi-structured interviews to examine verification practices, refusal experiences, and workflow fit. This paper makes two contributions:

\begin{enumerate}
    
    \item \textit{Empirical:} We report findings from a five-month, in the wild, mixed-methods deployment of AVA, engaging over 2,200 policy and development professionals across 116 countries. Triangulating longitudinal platform logs and pre–post surveys with 20 in-depth qualitative interviews, we characterize how these users integrate AVA into high-stakes workflows, detailing how they interpret reasoned abstentions as signals of system scope and trustworthiness and how they leverage page-anchored citations to support verification-driven evidence practices.
    
    \item[(2)] \textit{Conceptual}: We derive generalisable lessons for building trustworthy knowledge systems that can inform deployments in other high-stakes domains, including the need for an end-to-end trust pipeline (from corpus curation through abstention to verification), strategies for managing corpus quality–coverage trade-offs, and interface patterns that prioritise verification over disclosure. We also articulate a vision for ecosystem-aware specialised AI systems that prioritise collaborative interoperability with general-purpose models.
\end{enumerate}

\section{Related Work}

\noindent Our review synthesizes three strands: (i) epistemic humility and uncertainty communication, (ii) practice-grounded evaluation, and (iii) domain-bounded generative systems for specialized knowledge work. This lens motivates AVA’s design (reasoned abstention; verifiable citation) and our mixed-methods, in-the-wild evaluation, and situates AVA as a domain-bounded generative system built for knowledge professionals in evidence-intensive policy and development work.

\vspace{-10pt}
\subsection{Epistemic Humility and Uncertainty Communication}
HCI and Human Factors research has long framed appropriate reliance on automation as an interaction design problem: trust is better calibrated when systems make their limits as legible as their capabilities \citep{lee2004trust}. Foundational work shows how over- and under-reliance emerge, and argues for interfaces that support calibrated trust under uncertainty, motivating interface strategies such as confidence cues, boundary disclosures, and expectation-setting. For example, \citet{zhang2020effect} demonstrate that confidence signals and local rationales can modulate reliance in decision support, while \citet{kocielnik2019imperfect} show that acknowledging imperfection helps establish realistic mental models and temper over-trust. These findings frame trust calibration as a communication and interaction challenge: users must be able to see where a system ends, not only what it can do.

Recent studies shift attention from what a model says to how it refuses. \citet{wester2024denials} find that bare denials are rated more frustrating and less appropriate or useful than refusals that motivate or redirect. \citet{kim2024uncertainty} show that hedged uncertainty (e.g., “I am not sure, but…”) can reduce over-reliance and improve accuracy. Together, these findings support using reasoned refusals that briefly justify limits and offer a next step, rather than bare denials. In parallel, ML/NLP communities develop mechanism-level toolkits for abstention and grounding. For example, \citet{geifman2017selective} operationalize selective prediction for deep classifiers (a reject option under low confidence). More recently, LLM research advances evidencing and refusal: \citet{menick2022verifiedquotes} teach models to support answers with verified quotes, and \citet{gao2023alce} evaluate citation quality, finding many links only partially supportive, underscoring the difficulty of dependable evidencing. In a complementary line, \citet{chen2024teaching} align refusals so models decline or explain unknowns within bounded scopes.

These technical and interactional approaches converge on the concept of epistemic humility, which we synthesize across three traditions. Philosophically, epistemic humility is defined as an intellectual virtue grounded in recognizing one’s own fallibility and the limits of one’s knowledge~\cite{whitcomb2017intellectual,potter2022virtue}.
In HCI and CSCW, epistemic humility has recently been taken up as a design concern in multiple ways. Dedeoğlu et al.~\cite{dedeouglu2025navigating} frame it as a post-Enlightenment design value that resists technological mastery and foregrounds situated, plural forms of knowledge. Karusala et al.~\cite{karusala2024understanding} similarly argue that public-sector algorithms must practice humility to avoid over-claiming what they know about human subjects, particularly in sensitive decision-making contexts. Within the ACM community, Knowles et al.~\cite{knowles2023humble} crystallize this stance under Humble AI, a design approach emphasizing scepticism (acknowledging the limits of statistical proxies and resisting over‑confident claims), curiosity (interrogating unknowns by seeking evidence of “trust‑responsiveness,” especially in borderline or rejected cases), and commitment (avoiding distrust of the trustworthy, even at the expense of efficiency, by prioritising fairness, inclusion, and ongoing responsible practice).

Building on this lineage, we operationalise epistemic humility for generative systems as a system capability: the ability to make knowledge boundaries explicit, to withhold certainty when evidence is insufficient, and to prioritise verifiable, evidence-backed responses over plausible but unsupported guesses. While empirical work in domains such as algorithmic hiring has demonstrated how surfacing algorithmic unknowns and communicating uncertainty can support calibrated reliance in high-stakes workflows~\citep{nair2025humble}, much of this scholarship remains either conceptual or anchored in narrow-domain deployments, leaving open how epistemic humility can be operationalised in generative systems used at scale for professional knowledge work.

AVA addresses this gap by operationalising epistemic humility through two complementary interaction mechanisms in a domain-bounded generative platform. First, reasoned abstention declines unsupported queries with a brief justification and constructive redirection, making scope boundaries explicit and reducing unsupported generation. Second, verifiable citation attaches page-level references to factual statements, enabling users to independently check and contest claims. In combination, these mechanisms operationalise humility as a user-facing property that supports verification and accountable use.

\subsection{Evaluation in Practice}
Recent position work argues that GenAI evaluation must be sociotechnically grounded and practice-based: metrics should be iteratively refined from what people actually do with systems in situ, and studies should report design and measurement choices transparently \cite{weidinger2025evalscience}. Complementary frameworks call for in-the-wild, lifecycle-oriented evaluation that uses dynamic, outcome-oriented measures (beyond accuracy), combining deployment traces with user research and organizational context to capture value and risk in real work \cite{jabbour2025wild}. A recent systematic review of 153 CHI papers (2020--2024) highlights persistent validity and reproducibility issues, a predominance of artifact-centric, short-horizon studies, and a relative absence of multi-month deployments \cite{pang2025llmification}. While task-supportive user studies suggest that adapted interaction scaffolds can reduce cognitive burden in professional settings, such evaluations are typically scenario-based and short-duration \cite{wang2024task}.

Guided by this agenda, we examine axes that matter for real adoption in professional settings. We analyze what users actually do with verification-oriented affordances (page-level citations; reasoned refusals with redirection) alongside other adoption-relevant axes: task fit, adoption dynamics, workflow integration, multilingual/domain coverage, thus operationalizing calls to link measurement to situated activity (e.g., session flows, handoffs, reformulations) \cite{weidinger2025evalscience,jabbour2025wild}. With AVA, we address two gaps in the literature:
\begin{enumerate}
  \item \textbf{Longitudinal, deployment-scale evidence.} Despite calls for practice-grounded assessment, there are few multi-month, mixed-methods deployments in evidence-dependent professional domains that connect log-level behaviors with self-reported outcomes and workflow narratives \cite{weidinger2025evalscience,pang2025llmification}.
  \item \textbf{Multilingual, domain-bounded practice.} Methods for evaluating multilingual coverage and workflow integration in domain-bounded tools used by globally distributed professionals remain limited; most existing evidence is short-horizon or scenario-based \cite{jabbour2025wild,pang2025llmification,wang2024task}.
\end{enumerate}

We report a five-month, mixed-methods deployment of AVA with 2{,}200{+} users across 116 countries. Consistent with practice-grounded evaluation, we analyze real-world use across multiple axes, including task fit, adoption dynamics, workflow integration, multilingual/domain coverage, verification and contestability, learning curves, and boundary conditions, using platform logs, surveys, and interviews.

\subsection{Generative AI for Specialized Knowledge Work}

Generative research tools for knowledge work optimize differently for scale, trust, and relevance, often leaving the evidentiary needs of policy and development practice only partially served. While the HCI community has developed domain-specific systems aligned with professionals’ epistemic norms and workflows~\cite{karnatak2025acai,Karnatak2025ExpandingGenAIDesignSpace,fok2024marco}, prior efforts largely target contexts that differ from the institutional and evidentiary demands of international development and policy research.
Accordingly, we focus on a class of generative research agents designed to support information retrieval and synthesis, and examine how their design trade-offs create systematic epistemic gaps for policy and development work. We categorize these tools into three broad classes to highlight the specific epistemic gaps that AVA addresses.

\noindent
First, open-web and “deep research” agents (e.g., Perplexity) prioritize scale by drawing on the public internet~\cite{perplexity2024answerengine}. While providing broad coverage, these systems can inherit ranking and visibility biases from web search and attention dynamics, which may over-surface highly visible public content (e.g., news and summaries) relative to dense operational or grey-literature reports that often underpin policy decisions~\cite{fortunato2006topical}.

Second, user-defined retrieval systems (e.g., Google NotebookLM) originally prioritized trust by restricting generation to user-supplied documents~\cite{GoogleNotebookLMHelp2025}. While recent updates have introduced "Deep Research" agents to autonomously scour the web for literature, transitioning from this closed-loop system to open-web retrieval presents a distinct trade-off: it alleviates the bottleneck of user knowledge but re-introduces the challenge of verifying the provenance of autonomously selected sources.

Third, academic discovery engines (e.g., Elicit, Emergent Mind) optimize for scale and trust within peer-reviewed literature~\cite{ElicitAI2025}. However, these tools face a problem of structural invisibility regarding development data. Academic agents rely on structured metadata, such as Digital Object Identifiers (DOIs) and citation graphs, to traverse the web. Grey literature including field evaluations, policy notes, and operational reports typically lacks these structural features~\cite{Schopfel2010Grey}. Consequently, valuable operational knowledge remains in the "hidden web," effectively invisible to agents that rely on bibliographic crawling.

To address these gaps, we introduce AVA, a domain-specific generative research agent for policy and development work. AVA is a multi-agent, domain-bounded RAG system built on a verified library of over 4,000 World Bank Reports. Unlike open-web agents that seek to find new information, AVA is designed to reliably synthesize trusted information. This approach leverages Retrieval-Augmented Generation (RAG) to restrict the model's non-parametric memory to a curated corpus~\cite{lewis2020retrieval}. By bounding retrieval, AVA addresses the ranking bias of open agents and the structural blindness of academic agents, centering the grey literature that policy professionals routinely use~\cite{Schopfel2010Grey}. Table \ref{tab:ava-comparison} situates AVA within this design space. We contribute an empirical analysis of how policy and development professionals across 116 countries appropriate this domain-bounded assistant in their everyday evidence-based workflows.

\vspace{-10pt}

\section{System Architecture}
\noindent\textbf{Overview.}

AVA is a transparent, multilingual, multi-agent, evidence-grounded RAG system that provides accurate, verifiable, and context-aware answers. The architecture pairs a hierarchically indexed, expert-curated corpus of 4000+ World Bank Reports with agentic, tree-aware retrieval and an ensemble framework that ensures accurate citations and abstention. We next formalize the design goals and resulting architectural workflow.

\begin{figure*}
  \centering
  \includegraphics[width=\linewidth]{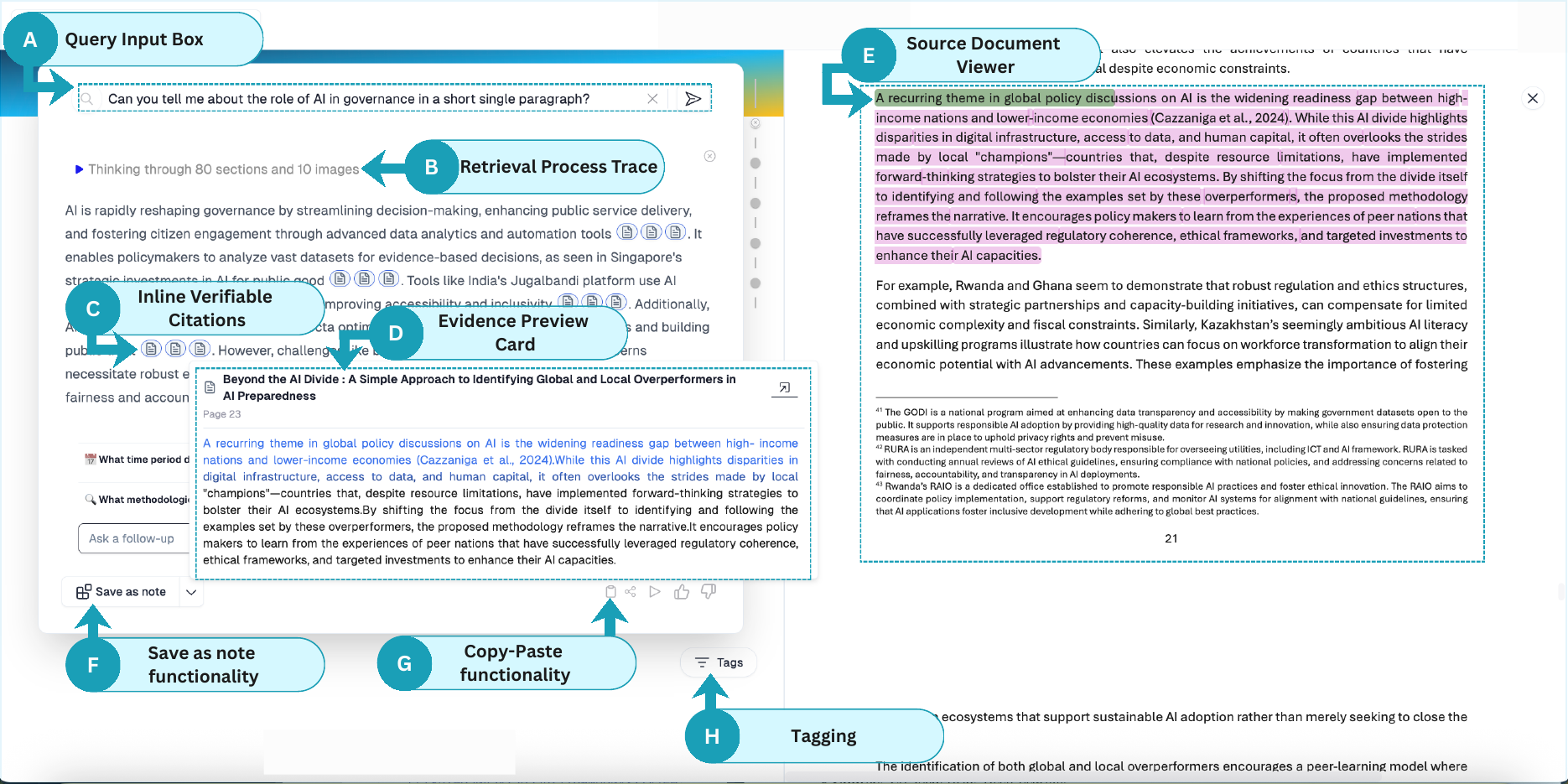}
  \caption{AVA interface annotated with its core components: (A) query input box, (B) retrieval process trace, (C) inline verifiable citations, (D) evidence preview card, (E) source document viewer, (F) Save as note functionality, (G) Copy-Paste functionality, and (H) Tagging. See appendix \ref{app:walkthrough} for detailed descriptions of each component.}
  \Description{A five-step user flow in AVA: (1) submit a query; (2) check the returned citations; 
  (3) verify that sources are credible and relevant; (4) save the verified response to notes; 
  (5) edit the response if needed, with arrows indicating a loop between edit and save. 
  Screenshots illustrate each step and the emphasis on citation-anchored verification.}
  \label{fig:ava-walkthrough}
\end{figure*}

\subsection{Design Goals}
The system architecture is driven by four primary design goals:

\noindent\textbf{DG1 - Verifiability and Epistemic Humility}
Every generated claim must trace to specific source spans, with the system preferring reasoned abstention over unsupported assertions. This epistemic humility, addresses a core challenge in knowledge-intensive AI systems. Prior works on RAG-based systems demonstrate that accountable generation depends on hybrid retrieval combining sparse lexical matching with semantic embeddings, enabling claims to bind to verifiable evidence while maintaining high recall across paraphrase and language variation \cite{zhang2025mixtureofragintegratingtexttables}. We operationalize this through hierarchical document indexing in the RAG and mandatory source attribution and confidence thresholds that trigger non-answers when evidence is insufficient or conflicting.

\noindent\textbf{DG2 - Workflow Fit} Effective human–AI collaboration in knowledge work requires inspectability and reproducibility. AVA therefore provides citation-first outputs, exportable artifacts, and version-pinned runs that support fluent, audit-ready workflows.

\noindent\textbf{DG3 - Multilingual Access}
Retrieve over original-language materials while responding in the user's language, treating parity of support across languages as a first-class usability criterion. Real-world policy and development research spans multilingual sources, yet translation introduces systematic errors that compound retrieval limitations. We extend RAG's grounding promise to multilingual settings by preserving original citations while ensuring response quality. This is a core requirement for equitable global deployment, designed to serve a diverse community in their native languages.

\noindent\textbf{DG4 - Responsible Personalization}
Enable user-controlled personalization through explicit preference settings that steer retrieval and ranking while maintaining evidence standards and user agency. Effective personalization in socio-technical contexts draws on groun-ded signals rather than opaque heuristics. We constrain personalization to opt-in predicates that users can inspect and modify, expose when preferences affected results, and preserve user control over system memory. This approach parallels recent work using interaction traces to align AI behavior with group norms while preserving existing practices and user autonomy.

\subsection{Core Components}
We proceed module-by-module, moving from data foundations to retrieval, from retrieval to verification, and finally to user-facing transparency and personalization. For additional details, see Appendix \ref{app:system_details}.

\subsubsection{Data Curation and Hierarchical Indexing}
\label{sec:stage1}
We transform over 4,000+ World Bank Reports from PDFs into a knowledge base using custom layout parsing and OCR. This extracts text, tables, and spatial relationships while preserving document hierarchy. Documents are chunked into nodes at paragraph  and cell levels. Each node receives a stable ID and hierarchical path (e.g., /section/paragraph). Nodes are bounded at 2048 tokens. This enables precise span-level citations, directly operationalizing DG1.

The processed data feeds into a dual-backend index: A graph store captures the  hierarchical tree, while a vector index holds  semantic embeddings from Qwen3-Embedding-8B. This setup supports structural traversal and cross-lingual search to meet DG3.

\subsubsection{Agentic Retrieval and Evidence Formulation}
\label{sec:stage2}
A multi-agent architecture decomposes complex queries. This mitigates single-agent ReAct pitfalls (role drift, unstable planning) \cite{yao2023react,shinn2023reflexion,li2023camel} by assigning distinct competencies:\\
The \textbf{Query Decomposer Agent} extracts atomic sub-questions, intents (factual, analytical, comparative), and targets. Decomposition improves accuracy and reduces hallucinations \cite{petcu2025querydecompositionragbalancing, ammann-etal-2025-question}. \\
The \textbf{Retrieval Planner Agent} selects search strategies (lexical vs. semantic). This hybrid planning outperform semantic or lexical retrieval \cite{sidiropoulos-etal-2021-combining}.\\
The \textbf{Tree-Walker Agent} traverses hierarchical structures and semantic neighborhoods. It switches between logical navigation (sections, references) and semantic exploration based on marginal gain, using coverage thresholds to stop. Graph-based traversal  surface more diverse evidence than flat retrieval \cite{zhang2025surveygraphretrievalaugmentedgeneration}.\\
Finally, the \textbf{Drafting Agent} consolidates retrieval paths into structured evidence packets containing relevant passages,  hierarchical contexts, and metadata. These undergo de-duplication and ranking (using GPT-4o-mini re-scoring combined with base retrieval) to present relevant evidence while preserving conflicting perspectives. This ensures inspectable outputs (DG2) and aligns with prior work on grounding stability~\cite{shao2025groundingaiexplanationsexperience}.

\subsubsection{Information Synthesis and Verification}
\label{sec:stage3}
Information synthesis employs a ``model orchestra" approach where specialized components handle distinct aspects of response generation before final integration. The information synthesis happens through the following four models: \\
\textbf{Evidence Classifier Model} (based on fine-tuned BERT models): This model performs passage classification, entity extraction, and span detection to identify key information within evidence packets.\\
 \textbf{Query Analyzer Model} (based on a fine-tuned GPT-4o-mini): This model handles query understanding, evidence summarization, and retrieval planning tasks that require reasoning but not extensive generation. \\
\textbf{Response Drafter Model} (based on a fine-tuned GPT-4.1 model): This model performs the primary response generation with structured, trace-based prompting that explicitly maps each generated claim to specific evidence spans within the packets. The prompting framework includes role specifications, evidence contextualization, and citation formatting instructions that ensure every factual assertion can be traced to verifiable sources.  \\
\textbf{Verification Model} (based on a fine-tuned GPT-4o-mini): Each draft answer is routed to a specialized verification model fine-tuned for fact-checking and inspecting tasks. The agent inspects sentence-level claims, aligns them with the retrieved evidence, and computes two internal scores:\\
(i) \textbf{Coverage}: the proportion of claims that have direct documentary support, and \\
(ii) \textbf{Agreement}: the degree of consistency across independent sources. The agent also judges whether the available evidence is rich enough to sustain a thorough explanation. \\
If either score falls below the high-confidence threshold i.e. confidence=high, or if the evidence is too sparse or contradictory for a complete response, the agent instructs the system to abstain rather than present potentially unsupported information. \\
This verification architecture directly operationalizes DG1's epistemic modesty principle. The system prefers abstention over confident but unfounded assertions, providing users with clear explanations of evidential limitations or insufficient coverage that prevented confident response generation. (See Appendix \ref{app:verification})

\subsubsection{User Personalization and Memory}
\label{sec:stage4}
The interface renders responses in users' preferred languages while preserving original-language citations and evidence spans, supporting DG3's multilingual access goals without introducing translation artifacts that could compromise verification. Interactive citations provide hover previews displaying source quotes in their original context and click-through links to highlighted passages in full documents. AVA currently supports queries in over 60 languages (see Appendix \ref{app:multilingual} for details).
 
The system maintains user-controlled memory through explicit save actions, allowing researchers to build personal knowledge collections, bookmark significant findings, and track research threads across sessions. All interactions are version-pinned with full citation trails, enabling users to export reproducible research artifacts, share findings with collaborators, and audit their analytical processes. When abstention occurs, the interface provides clear explanations of evidential gaps, suggests productive query refinements based on available evidence patterns, and offers related topics where stronger evidence exists, supporting iterative research workflows and collaborative analysis practices.

\section{Methodology}
\subsection{Study Design and Setting}
\begin{figure}
    \centering
    \includegraphics[width=\linewidth]{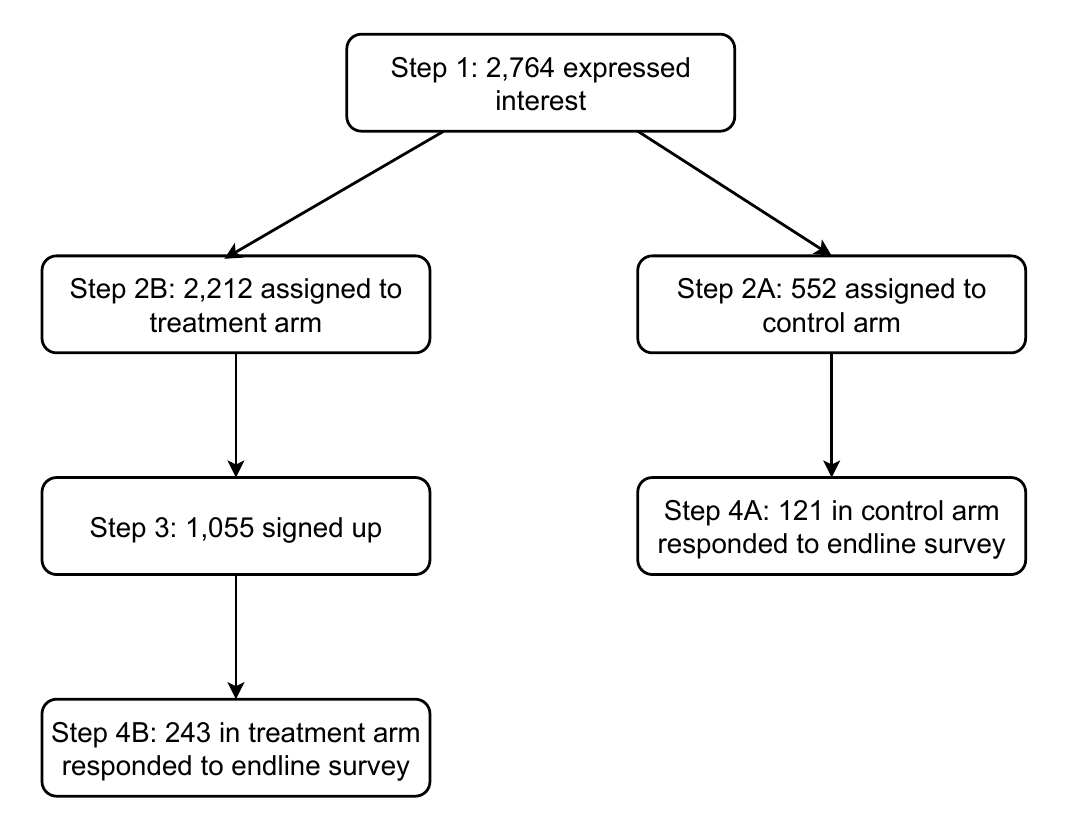}
    \caption{Participant flow. From 2,764 initial registrants, participants were assigned to treatment (n=2,212) and control (n=552) arms. The final analysis included 243 treatment and 121 control participants who completed the endline survey.} 
    \Description{Flowchart of participant progression and attrition in the study. From an initial 2,764 interested individuals, the flow splits into a control arm with 552 participants and a treatment arm with 2,212 participants. The chart concludes with the number of participants who responded to the endline survey: 121 from the control arm and 243 from the treatment arm.}
    \label{fig:flowchart}
\end{figure}

We conducted a mixed-methods, randomized-invitation field evaluation of AVA. Participants were recruited via a large multilateral development bank (MDB). Employment at the MDB was not required; the sample comprised predominantly external practitioners working in policy, development, and related fields. Specifically, only 4.6\% of registered participants were from the host institution; the remaining 95.4\% were external participants. Participation was voluntary and secured via informed consent. Following baseline collection, registrants were randomized 80:20 into a treatment arm (offered immediate access) and a holdout control arm (no access). We employed an open-label intention-to-treat (ITT) design, analyzing outcomes via difference-in-differences (DiD) estimation to account for pre-existing differences (see Figure ~\ref{fig:flowchart} for participant flow). The evaluation window extended through August 31. Detailed recruitment procedures, timeline milestones, and author positionality are provided in Appendix \ref{app:methodology_extended}

\subsection{Quantitative Methods}
\subsubsection{Data Sources and Measures}
We integrated four sources: (i) baseline and endline surveys (demographics, roles, prior AI usage, adoption, productivity, satisfaction); (ii) in-app pop-up surveys during deployment (trustworthiness, recommendation, satisfaction); (iii) interaction logs (timestamps; session identifiers; query text; language; response metadata including citation counts; per-response feedback); and (iv) semi-structured interviews with users. Survey records were linked to platform accounts via the email used at registration; operational logs were pseudonymized with hashed identifiers.
We operationalized key constructs  including (1) AI Use for Work, (2) AI-Driven Productivity Gains, and (3) Perceptions of AVA. Detailed items and scoring rules are provided in Appendix \ref{app:methodology_extended}.

\subsubsection{Quantitative Data Analysis} \hfill\\

\textbf{Query Classification:} We applied natural language processing (NLP) to classify user queries into policy themes and intent categories. The pipeline first normalized and sessionized text, automatically translating non-English queries into English. We then employed a deterministic, rule-based classification system using curated taxonomies to map queries to specific domains (e.g., Health, Infrastructure) and interaction types. To ensure coverage, uncategorized queries were imputed based on session context (forward-filling within a one-hour inactivity window). Detailed preprocessing steps, matching logic, and taxonomy definitions are provided in Appendix \ref{app:methodology_extended}.

\textbf{Behavioral Metric Derivation:} Interaction logs were processed to derive the core behavioral metrics reported in Sections 5.1–5.3. We calculated Abstention Rates by tracking the proportion of queries where the system triggered a refusal state due to insufficient context (see Fig. ~\ref{fig:5_day_total_queryvolume}). Citation Density was computed by extracting the count of unique document node references per generated response to quantify evidence reliance (reported in Section 5.1). Finally, to analyze Sustained Engagement, we segmented users into single-session and multi-session cohorts based on the session grouping logic; these cohorts serve as the basis for the user retention analysis (Section 5.2) and the Difference-in-Differences efficiency estimates (Section 5.3)

\textbf{Regression Analysis. }
To estimate causal effects of access to AVA, we conducted intention-to-treat (ITT) analyses comparing treatment and control groups using endline survey outcomes, complemented by difference-in-differences (DiD) estimation that leveraged both baseline and endline responses. Linear regression with robust standard errors was used throughout. Key self-reported outcomes included productivity (e.g., time saved, number of outputs), AI tool adoption frequency, and perceived work quality.

\subsection{Qualitative Methods}
\subsubsection{Sampling and Participants} From survey respondents who consented to follow-up, we used purposive sampling to achieve variation in geography, role and language. We conducted $n=\,[$20$]$ semi-structured interviews via secure video calls. Table~\ref{tab:participant-demographics} reports participants' demograhic details. Our participants included 8 women and 12 men; age bands were 19–29 ($n=3$), 30–39 ($n=10$), 40–49 ($n=4$), and 50–59 ($n=3$). Career stage spanned early ($n=9$), middle ($n=7$), and senior ($n=4$) professionals across sectors (universities, government, NGOs/civil society, international organizations, startups, private sector). Participants represented multiple regions, with substantial coverage from Africa ($n=11$ across Nigeria, Uganda, Ghana, Tanzania, Angola), as well as the Americas (USA, Argentina, Canada), Europe (France), and Asia (Malaysia). Linguistic diversity was high (11 reported multilingual proficiency, e.g., English with Luo/Swahili, Igbo, Yoruba, French, Spanish, Mandarin, Arabic).

\subsubsection{Data collection}
Data was collected through semi-structured interviews conducted via secure video calls. Each interview lasted 45--75 minutes. The interview protocol guided conversations to explore, users' work and task contexts, perceived benefits and risks of using AVA, strategies for trust calibration (\textit{e.g.,} use of citations and handling of abstentions), effective prompting techniques and cross-tool comparisons. All interviews were audio-recorded with participant consent, transcribed verbatim, and then de-identified to ensure confidentiality.

\subsubsection{Data Analysis} 
The first author conducted all interviews and prepared verbatim, de-identified transcripts. We employed thematic analysis. An initial codebook was developed from an open-coding pass on a subset of transcripts by the first author and circulated to the team for critique. The team independently reviewed the codebook, proposed refinements by jointly coding a common subset. Discrepancies were resolved through discussion; the refined codebook then guided subsequent analysis. The first author completed coding of the remaining transcripts and led synthesis, with regular team debriefs. We combined deductive codes aligned with our research questions (e.g., evidence use, abstention handling, prompting challenges) with inductive codes grounded in the data. Recruitment and interviewing stopped once meaning saturation was reached.

\subsection{Mixed-Methods Integration (Triangulation)}
We integrated quantitative and qualitative strands at the interpretation stage. Survey and causal estimates characterize population-level patterns, while interaction logs provide granular behavioral traces. Interviews explain mechanisms underlying these patterns (e.g., why page-anchored citations serve as evidence shortcuts, how abstention calibrates reliance). We report convergences and divergences across strands and link them to design implications.

\noindent
In Sections 5, we first present quantitative analyses that characterize broad engagement patterns and their associations with self-reported impacts. We then draw on qualitative data, in section 6, from 20 semi-structured interviews with a with a diverse, global cohort of users, including university lecturers, independent researchers, NGO professionals, and startup founders (see Table ~\ref{tab:participant-demographics}), to explain the contexts and mechanisms underlying these patterns.

\section{Quantitative Results}
\subsection{Product Evaluation: How Users Engage with AVA (RQ1)}

\begin{figure*}
    \centering
    \includegraphics[width=0.7\linewidth]{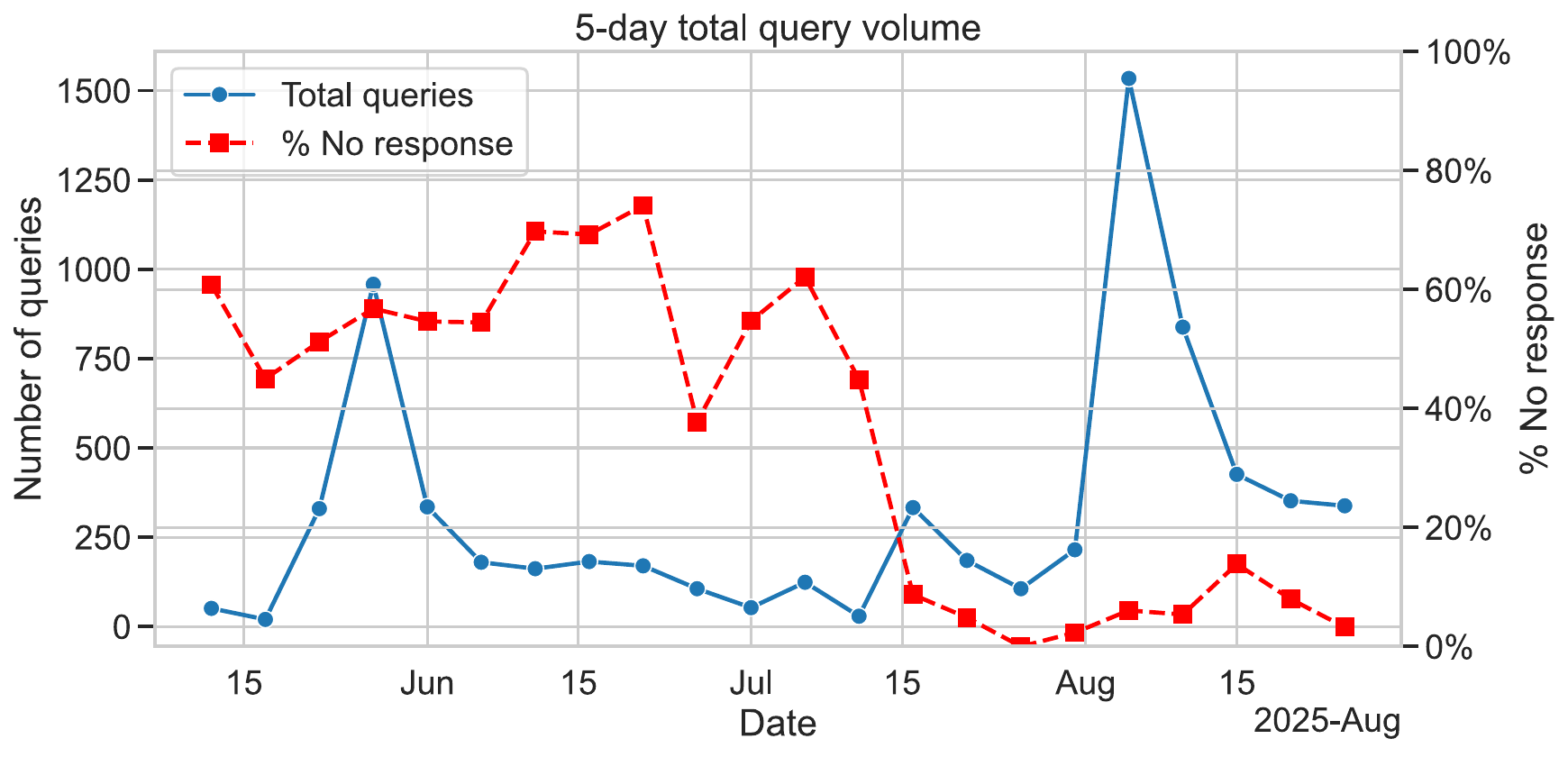}
    \caption{Five-day total query volume and abstention rate (\% of queries receiving a reasoned “no response”). Abstention is high during early deployment with limited corpus coverage and drops sharply after corpus expansion, remaining low thereafter.}
    \Description{A dual-axis line graph showing the 5-day total query volume and the percentage of no-response queries from May to August 2025. The number of total queries (left axis) shows two significant peaks: one in early June at approximately 1,200 queries and a sharper peak in early August reaching nearly 1,600. The percentage of queries with no response (abstention rate) (right axis) generally follows an inverse trend, dropping to its lowest point of near zero percent during the major query peak in August.}
    \label{fig:5_day_total_queryvolume}
\end{figure*}

During the study period, from May 12 to August 31, 2025, 2,679
unique individuals signed up for AVA. Of these, 1,055 users (39.4\%) were successfully matched to the baseline survey via email addresses. Among the 1,037 matched users with country information, 782 (75.4\%) were based in the Global South, including Nigeria, India, and Brazil (see Fig \ref{fig:global_dist}). Although users were geographically diverse, most queries were in English (83.8\%), followed by French (4.8\%) and Spanish (3.0\%).
Figure~\ref{fig:5_day_total_queryvolume}  shows the total volume of queries over time alongside the proportion of queries AVA declined to answer due to insufficient corpus coverage. Early in the deployment, when the knowledge base contained approximately 50 reports, the abstention rate ranged from 40–70\%. Following the expansion of the corpus to over 4,000 World Bank reports on 9 July 2025, the abstention rate dropped sharply and remained below 10\% for the remainder of the study period. Across the full deployment, responses generated by AVA cited an average of five documents per query (range: 1–23), indicating consistent reliance on multi-document evidence.\\

\noindent
To understand how users engaged with AVA, we first classified queries by policy theme (See Table~\ref{tab:policy-themes}). Two themes, namely, Human Capital and Fiscal Policy/Private Sector dominated, accounting for roughly two-thirds of all queries. These queries focused on issues such as jobs, education, poverty, growth, debt, and taxation trade-offs, indicating strong demand for evidence-informed policy synthesis. Example questions were largely how-to and forward-looking (e.g., “How should labor regulations evolve post-pandemic?”; “How can policymakers balance investment and innovation?”), suggesting that users sought support in framing complex policy questions and identifying plausible policy directions, rather than retrieving narrow factual information.
We further classified queries by primary task type—diagnostics, design, and evaluation (See Table~\ref{tab:summary_Of_Policy_Tasks_Reflected_in_Queries}, Fig \ref{fig:ques_themes}). Our analysis differentiates between queries focused on diagnostics (seeking to understand problems, data, and challenges), design (how to improve policies, find best practices), and evaluation (evidence, effectiveness measures, and impact evaluation support). Overall, we find that most queries ask AVA to complete diagnostic tasks (See Table~\ref{tab:summary_Of_Policy_Tasks_Reflected_in_Queries}, Fig \ref{fig:ques_themes}), although the content of queries reinforce the use of AVA throughout the policymaking process,from defining problems, designig solutions, and testing impact.

\begin{table*}[htbp]
\centering
\caption{Summary of policy themes reflected in queries}
\Description{This table summarizes the main policy themes reflected in user queries. It's organized into five categories: Human Capital, Fiscal Policy, Digital Transformation, Environment, and Infrastructure. For each theme, the table provides the top keywords, an example query, and a corresponding value indicating its frequency.}
\label{tab:policy-themes}
\resizebox{\textwidth}{!}{%
\begin{tabular}{|l|l|c|p{0.45\linewidth}|}
\hline
\textbf{Policy Theme} & \textbf{Top Keywords} & \textbf{\% of Queries} & \textbf{Example Query} \\
\hline
Human Capital & education, school, poverty, program, job & 33.5 & How should labor regulations evolve post-pandemic? \\
Macro-Economics and Governance & government, income, growth, economy, taxation & 20.7 & How do countries balance investment and innovation? \\
Digital Transformation & data source, tools, technology used, analyze, analytic & 17.4 & What types of digital infrastructure hinder ing digital transformation in Nigeria and Kenya? \\
Environment and Climate & fish, waste, emissions, air, age & 13.5 & How to integrate adaptation and mitigation for climate change in Nigeria? \\
Infrastructure & cargo, earthing, clearance, cargo service, circuit & 5.7 & How to amplify non-motorized transport for sustainable urban mobility in Uganda? Highlight indigenous paradigm. \\
\hline
\end{tabular}%
}
\end{table*}

\begin{table*}[t]
\centering
\caption{Summary of policy tasks reflected in queries.}
\label{tab:summary_Of_Policy_Tasks_Reflected_in_Queries}
\resizebox{\textwidth}{!}{%
\begin{tabular}{|l|l|c|p{0.45\linewidth}|}
\hline
\textbf{Policy Task} & \textbf{Sample Keywords} & \textbf{\% of Queries} & \textbf{Example Query} \\ \hline

Diagnostic &
what, explain, data on, impact of, challenges &
69.0 &
What are measures used in assisting Young Africans and changing their mindset? \\ \hline

Design &
how to, improve, best practice, options for, recommendation for &
22.0 &
Write a proposal for masters degree thesis on how to make Ghana government better. \\ \hline

Evaluation &
evaluation, assessment, evidence, result, effectiveness &
9.0 &
I am conducting an impact evaluation on an LPG provision scheme. Please identify potential proxy variables for income. \\ \hline

\end{tabular}%
}
\end{table*}

\noindent
Notably, evidence from our endline survey suggests that use of AVA may have substituted use of other AI products. When reporting the AI services they used in the last week, only 24\% of AVA users, those who were invited to test the product and actually used it, reported also using another AI service (e.g., ChatGPT, Gemini, etc.) whereas 80\% of non-AVA users (participants who signed up for AVA but were not yet given access) reported using another AI service.  
\noindent

\subsection{User Evaluation: Satisfaction With and Perceptions of AVA (RQ3)}
\noindent
As a behavioral indicator of satisfaction, we explored the proportion of users who engaged with AVA in a single session or multiple sessions and characteristics associated with engagement in multiple sessions. Overall, we found that 73.4\% of total AVA users engaged in only one session whereas 26.6\% of users returned to the platform to engage in multiple sessions. While most users did not return for multiple sessions, the rate is comparable or better than industry benchmarks for productivity apps (\href{https://www.statista.com/statistics/259329/ios-and-android-app-user-retention-rate/}{Statista, 2025}). In Table~\ref{tab:Summary_statistics_of_characteristics_of_registered_users_and_users_with}, we also summarize the characteristics of baseline respondents who engaged with AVA for a single session or multiple sessions, finding that users had overall similar profiles.

\begin{table}[t]
\centering
\footnotesize
\caption{Summary statistics of characteristics of registered users and users with different levels of engagement.}
\Description{This table provides a statistical summary comparing the baseline characteristics of three distinct user groups: all Registered users, users who had only a Single Session, and highly engaged users with Multiple Sessions.}
\label{tab:Summary_statistics_of_characteristics_of_registered_users_and_users_with}
\resizebox{\linewidth}{!}{
\begin{tabular}{|p{0.45\linewidth}|c|c|c|}
\hline
& \makecell{\textbf{Registered}\\ \footnotesize N = 1,055}
& \makecell{\textbf{Single Session}\\ \footnotesize N = 506}
& \makecell{\textbf{Multiple Sessions}\\ \footnotesize N = 254} \\
\hline
\textbf{Global South (\%)} & 74.2 & 74.1 & 74.4 \\
\hline
\multicolumn{4}{|l|}{\textit{Affiliation (\%)}} \\
\hline
\quad Think Tank                 & 4.2 & 4.3 & 5.5 \\
\quad Academic                   & 15.7 & 15.8 & 15.7 \\
\quad Multilateral Organizations & 6.0 & 6.3 & 3.9 \\
\hline
\multicolumn{4}{|l|}{\textit{AI Experience}} \\
\hline
\quad Multiple Times Every Day                 & 39.1 & 38.3 & 44.9 \\
\quad About once or twice on most days         & 18.8 & 20.2 & 14.2 \\
\quad A few times throughout the week          & 25.7 & 25.3 & 26.4 \\
\quad Rarely                                   & 5.0  & 4.7  & 3.5  \\
\quad Not at all                               & 1.5  & 1.4  & 2.0  \\
\quad Never use AI for work                    & 9.9  & 10.1  & 9.1  \\
\hline
\end{tabular}
}
\end{table}

\noindent
Based on the endline perception data from 118 AVA users, the tool is viewed very positively. A high percentage of users found AVA to be an effective and reliable resource, with strong endorsement for its efficiency, accuracy, and overall value. 
In addition to the behavioral indicators of satisfaction, we also collected survey data from individuals using AVA as a pop-up survey to assess perceptions of the tool. Results from 100 individuals, indicated that 68\% find content relevant, 65\% the citations relevant and 72\% were satisfied with AVA and would recommend the tool to a colleague. 

\begin{table*}[t]
\centering
\footnotesize
\caption{User perceptions of AVA from the endline survey (n = 118). Percentages indicate respondents who agreed with each statement.} 
\label{tab:user_agreement} 
\begin{tabular}{|l|c|} 
\hline
 & \begin{tabular}[c]{@{}c@{}}Percentage of users \\ who agree with the \\ statement (n=118)\end{tabular} \\
\hline
AVA helps me understand key insights from World Bank documents more efficiently. & 80.5\% \\
\hline
The synthesized insights provided by AVA are accurate and trustworthy. & 82.2\% \\
\hline
Using AVA has improved the quality of my work or research. & 76.3\% \\
\hline
I would recommend AVA to others working in development policy or research. & 89.0\% \\
\hline
\end{tabular}
\end{table*}

\subsection{Impact Evaluation: Improvements to Work Efficiency (RQ3)}
To assess the impact of AVA, we asked respondents of the endline survey to rate how much time they have saved on work through the use of AI and the quality of their work. We first ran an intention-to-treat (ITT) analysis, assessing differences in time saved on work among endline respondents who were invited to register to use AVA and those who were assigned to the waitlist control. ITT analysis revealed no significant differences in time saved between the treatment and control groups for either the lower bound (b = -0.56, p > .05) or upper bound estimates of time saved (b = -0.78, p>.05).  In addition, no impacts were found when we asked respondents to estimate the number of major outputs they completed in the last week and report if the use of AI tools helped them produce a greater number (b = 0.08, p > .05) of outputs or saved time in producing those outputs (b~=~-0.78,~p~>~.05).

\begin{table}[t]
\centering
\footnotesize
\caption{Intention-to-treat (ITT) Regression Results} 
\Description{This table presents the results from four regression models designed to measure the impact of an 'Intent-to-Treat' (ITT) condition on four different productivity outcomes: time saved (lower and upper bounds) and the number of additional or faster outputs. The table displays the estimation coefficients and robust standard errors in parentheses.}
\label{tab:regression_results} 
\resizebox{\linewidth}{!}{
\begin{tabular}{|@{}|l|c|c|c|c|@{}|}
\hline
& (1) & (2) & (3) & (4) \\
\hline
VARIABLES & \begin{tabular}[c]{@{}c@{}}Time AI tools save \\ (lower bound) - \\ Hours\end{tabular} & \begin{tabular}[c]{@{}c@{}}Time AI tools save \\ (upper bound) - \\ Hours\end{tabular} & \begin{tabular}[c]{@{}c@{}}Additional \\ outputs produced \\ because of AI - \\ Number\end{tabular} & \begin{tabular}[c]{@{}c@{}}Outputs \\ completed faster \\ because of AI - \\ Number\end{tabular} \\
\hline
Intent-to-Treat & -0.568 & -0.777 & 0.0847 & -0.225 \\
& (0.416) & (0.742) & (0.334) & (0.382) \\
& & & & \\
Constant & 4.010$^{***}$ & 7.172$^{***}$ & 2.444$^{***}$ & 3.909$^{***}$ \\
& (0.341) & (0.616) & (0.277) & (0.305) \\
\hline
Observations & 289 & 289 & 288 & 289 \\
R-squared & 0.007 & 0.004 & 0.000 & 0.001 \\
\hline
\multicolumn{5}{l}{\footnotesize Robust standard errors in parentheses. $^{***}p<0.01$, $^{**}p<0.05$, $^{*}p<0.1$} \\
\end{tabular}
}
\end{table}
\noindent
In addition to ITT, we conducted a difference-in-difference (DiD) analysis to explore whether different intensity levels of AVA usage (single use vs. multiple session use) were associated with different levels of impact on work efficiency. Here, we find that individuals who engaged with AVA over multiple sessions reported saving more time on their work in the last week for both the lower (b = 1.84, p < .10) and upper bound of estimates (b = 3.27, p < .10). 

\begin{table}[t] 
\centering
\footnotesize
\caption{Difference-in-Differences Regression Results} 
\Description{This table displays the results from a Differences-in-Differences (DiD) analysis. This model is used to estimate the causal effect by comparing a AVA returning users with one time AVA users over time. The two models in the table estimate the intervention's impact on the lower (1) and upper (2) bounds of time saved, measured in hours. The table displays the estimation coefficients and robust standard errors in parentheses.}
\label{tab:did_results} 
\begin{tabular}{|@{}|l|c|c|@{}|}
\hline
& (1) & (2) \\
\hline
VARIABLES & \begin{tabular}[c]{@{}c@{}}Time AI tools save \\ (lower bound)\end{tabular} & \begin{tabular}[c]{@{}c@{}}Time AI tools save \\ (upper bound)\end{tabular} \\
\hline
Round = Endline & -2.766$^{***}$ & -4.796$^{***}$ \\
& (0.705) & (1.253) \\
& & \\
One Time vs Returning & 0.554 & 0.640 \\
& (0.789) & (1.492) \\
& & \\
DiD (Intensity) & \textbf{1.838$^{*}$} & \textbf{3.272$^{*}$} \\
& \textbf{(1.035)} & \textbf{(1.886)} \\
& & \\
Constant & 4.946$^{***}$ & 9.027$^{***}$ \\
& (0.598) & (1.122) \\
\hline
Observations & 160 & 160 \\
R-squared & 0.132 & 0.112 \\
\hline
\multicolumn{3}{l}{\footnotesize Robust standard errors in parentheses. $^{***}p<0.01$, $^{**}p<0.05$, $^{*}p<0.1$} \\
\end{tabular}
\end{table}


\noindent

\section{Qualitative Results}
We organize our qualitative findings into four key themes. First, we detail how participants integrated AVA into their daily evidence workflows, its distinctive value, and professional impact (Section 6.1). We then explore the mental model and persona they developed around its unique behavior, particularly reasoned abstention (Section 6.2). Next, we analyze the foundations of trust in AVA through both its features and institutional corpus (Section 6.3). Finally, we discuss the ethical debates on disclosure that emerged from its successful adoption (Section 6.4).\\

\subsection{Positioning AVA within Evidence-Based Workflows (RQ1)}

In this section, we report participants' integration of AVA into everyday evidence work. We analyze their strategic division of work across tools, their reliance on AVA's page-level citations, and their appreciation for its distinctive depth and policy-oriented synthesis, including graphics and visuals generated by synthesizing across documents. Finally, we demonstrate the translation of these capabilities into concrete professional achievements, enabling participants to prepare for high-stakes meetings, advance stalled projects, and publish with greater confidence.

\subsubsection{Multi-Tool Workflows and Ecosystem Strategies} Participants organized their work across complementary systems rather than relying on a single general-purpose model. Generalist LLMs (e.g., ChatGPT, Claude, Gemini) were used for ideation, outline scaffolding, and language polishing, while AVA was consistently positioned as the evidence engine for claim-level facts requiring citations and page-level grounding. As P1 explained: \textit{``I use ChatGPT to break down tasks when I am working on a project, and I use AVA to get the "real information" I need.''} Participants recognized AVA as a domain-specific tool grounded in World Bank Reports, turning to it for policy and development work. P2 noted: \textit{``Anything to do with policy, AVA is very straight, very direct and very helpful.''}

Participants also developed strategies for managing ecosystem boundaries. When AVA's curated corpus could not address queries, participants reported transitioning to other tools like Perplexity (P12). However, breakdowns with generalist tools, such as receiving false information, often prompted a return to AVA. As P9 shared,

\begin{quote}
\textit{Sometimes I use ChatGPT to do my analysis\ldots I say, `Please give me this, give me this,' and I discovered that one of the outputs that came out was false. So when I saw AVA, I said let me try this thing\ldots and lo and behold, correct information came out, a lot of references, a lot of case studies.}
\end{quote}

This preference for AVA extended beyond mere accuracy to the quality and nuance of its output. P18, who used ``different tools for different jobs,'' explained that while other models were useful for broad research, \textit{``\ldots the nuances of the data, Ava picks up a little better.''}. This orchestration demonstrates calibrated multi-tool strategies where participants developed quality hierarchies and task--tool matching based on empirical experience with different systems' strengths and limitations. 
\subsubsection{Citation Verification as an Evidence Shortcut (RQ3)}
The provision and verification of citations was consistently described as AVA's most valuable feature. Participants emphasized that citation verification supported their work through three key mechanisms. First, it accelerated source finding by narrowing thousands of pages of World Bank Reports, or even the internet, to only the most relevant sections. P6 noted: \textit{``Because it helped me streamline what I am looking for without wasting so much time searching for it on the web.''} Second, it facilitated scanning, enabling users to access highlighted excerpts in context rather than manually sifting through the entire documents.
Third, it streamlined fact checking, allowing participants to click citations and immediately confirm whether claims were accurately grounded in sources. These mechanisms produced massive efficiency gains. P9 quantified similar time savings: \textit{``Sometimes it takes me four days, and this AVA, maybe just five minutes to do my query, and in another one minute, I am getting the information I want. It's faster, easier.''} Collectively, these accounts highlight how AVA transformed citation verification from a resource-intensive bottleneck into a lightweight, reliable shortcut, simultaneously enhancing trust, narrowing reading scope, and increasing productivity.

\subsubsection{Distinctive Value Propositions of AVA} 
Based on their day-to-day use of AVA, participants contrasted it with general-purpose AI tools and highlighted four perceived advantages: deeper reference-rich grounding, improved readability, policy-oriented structuring, and visualisations that supported interpretation and communication in their workflows. Participants characterized AVA as a way to turn World Bank Reports into immediately usable professional insights. Several participants compared AVA with other AI tools and emphasized that AVA delivered deeper, reference-rich responses. As P9 highlighted:
\begin{quote}\itshape
``We are using other AI tools, but we are not getting what we want, and that's very deep knowledge and references. When I used AVA, it was deep. It gave me case studies, references.''
\end{quote}
\noindent
Participants also noted that AVA made World Bank materials easier to engage with. P13 described difficulty with statistics-heavy documents: \textit{``There are no stories in World Bank Reports, so people are not interested to read the statistics, but AVA makes it easier to read.''} Participants described AVA as moving beyond summary to policy synthesis; as P5 put it, \textit{``AVA tried to bring out some policy recommendations which you cannot see in some of the other AI platforms.''} P5 further noted that responses were organized into compact summaries and tables, helping them scan and locate relevant evidence quickly. His remark illustrates how participants differentiated AVA from the general-purpose AI tools they already used. While they knew other tools could generate generic ‘recommendations,’ participants treated AVA’s outputs as substantive policy recommendations because they were grounded in World Bank Reports and organised into compact, citable formats that made relevant evidence easy to scan and verify.
\noindent
Beyond text, participants shared that AVA's graphics (e.g., tables, charts, trends) helped them engage with World Bank Reports more effectively. For example, P14 shared
\begin{quote}\itshape
``When I was preparing to attend meetings at the United Nations (UN), I was able to get visualizations of information as opposed to having to take a data set and try to parse out key trends. I could just put in the prompt into the system and it auto-generated that information for me. In in terms of getting some quick stats I found it very useful in helping and assisting and amplifying my preparation.``

\end{quote}
\noindent
Across interviews, participants praised these visuals and requested the ability to export them as images with built-in attribution to AVA. These perspectives indicate that AVA surfaces and structures World Bank knowledge for professional use, supporting comprehension, evidence-building, policy articulation, and communication.

\subsubsection{Concrete Professional Milestones} AVA’s impact extended beyond efficiency to enabling professional confidence and substantive scholarly contributions. For example, P19, working on policy development for street children in Borno state, described using AVA to advance a book manuscript that had remained unsubmitted for years:  

\begin{quote}\itshape
``I have never submitted it to any superior before because I felt I needed more inputs. More feedbacks. Within some minutes, I was able to obtain positive feedback from AVA and I have submitted it to my superior.'' (P19)  
\end{quote}
\noindent
Other participants reported similar gains. P1 credited AVA- supported research with contributing to a successful publication accepted by an international journal, while P2 described progressing a book manuscript with AVA’s support.  
\noindent
Participants also noted that AVA facilitated preparation for both routine and high-stakes meetings. P20 used AVA to quickly ramp up on new development topics for team discussions, while P14 highlighted its role in preparing for United Nations meetings. 
\noindent
Finally, participants emphasized that AVA enhanced their confidence in professional interactions. As P9 reflected: \textit{``It gives me an extra lot of confidence because when I speak with some other colleagues who are also into development-related work\ldots{} there's this confidence I have right now that I'm going to give them very rich information until they ask further questions.''}  
\noindent
 These accounts suggest that AVA’s value extended beyond task completion to professional development and credibility, helping participants advance manuscripts to supervisory and scholarly review, prepare confidently for high-profile engagements, and speak authoritatively with peers.  

 \subsubsection{Auxiliary feature: Multilingual Capabilities}
Despite supporting 60{+} languages, AVA's multilingual capabilities were largely unused, 83.8\% of interactions occurred in English (see Fig.\ref{fig:top10}). When interviewed, participants (n{=}20) attributed this to workplace norms: they mostly conduct and publish research in English, despite knowing their native language. Those who tried  other languages rated quality highly; for example, P17 said \textit{``It's written in a good French, it's not a robotic French. It's pretty high level.''} Similarly, P7 reported high satisfaction with Spanish outputs.
Given the English-heavy usage patterns, our non-English sample remains small, limiting conclusions about multilingual performance.

\subsection{Reasoned Abstention: When AI Says ‘I Don’t Know’ (RQ2)}
\label{sec:abstention_findings}

In this section, we report participants' responses to AVA's reasoned abstention. We examine user reception and appropriation of "I don't know" responses as signals of honesty, evidence boundaries, and opportunities for prompt reformulation. We then describe the anthropomorphic interpretations that abstention fostered, with participants casting AVA as a reserved but trustworthy colleague while expressing aspirations for warmer, more supportive interaction. Finally, we present user suggestions for balancing epistemic honesty with workflow efficiency, including collaborative query refinement and external handoffs.
\subsubsection{Reasoned Abstention: Reception, Appropriation, and Limits} 
\label{sec:abstention} 
We designed AVA to practice reasoned abstention under \textsc{DG1}. The system states when evidence is insufficient, provides a brief justification, and offers targeted redirection. Participants generally welcomed this behavior in high-stakes work (e.g., policy drafting and development projects), preferring an explicit ``I do not know'' to speculative answers. P13 shared, \textit{``It does not fabricate, that's the credibility of AVA. It does not fabricate. If it doesn't know, it will not provide you data.''} This illustrates how some participants interpreted AVA’s abstention and citation mechanisms as strong safeguards against fabrication in this deployment. We examine how such trust is calibrated, and where it is contested, in Sections 6.3 and 7. Across participants, reasoned abstention was appropriated in multiple ways~\cite{dourish2003appropriation}, even though it was initially designed to prevent ungrounded claims.
For example, several treated abstention as a cue about evidence limits and a pointer to research gaps. P5 noted: \textit{``Even if AVA says I don't know, to me, it's very useful. `I don't know' is half of the answer to what I am looking for.''} P3 valued the honesty of non-response, while P7 contrasted AVA with general-purpose systems: \textit{``Sometimes they tend to give you an answer even when they are not pretty sure \ldots [that] gives you a misconception they will always have the answer.''} 

\noindent
Other participants treated abstention less as a definitive endpoint than as an invitation to refine their information-seeking approach. P6 described systematically rephrasing queries until achieving results, while P7 positioned abstention as a brainstorming catalyst. These users treated abstention as an invitation for prompt reformulation and negotiation.

\noindent
However, not all participants welcomed abstention. Some (P4 and P18) interpreted it as system limitation rather than calibrated caution. P18 shared: \textit{``I don't want the choice being no answer. If you keep saying I don't know, I'm just going to go and use Google.''} These dissenting views highlight the tension between abstention as a trust-building mechanism and abstention as a productivity barrier.  We discuss this in the following sections.

\noindent

\subsubsection{Two Modes of Anthropomorphism: Descriptive Personas and Aspirational Desires}

AVA's reasoned abstention created an unexpected pathway to anthropomorphization, helping participants make sense of its system boundary and develop workable mental models. Participant's connected AVA’s behavior to familiar human communication norms. As P19 explained: \textit{``For me as a human, whenever you ask me something that I don't know, I will be truthful and honest, and if I know places you can reach out to get that information, I would redirect you there. Incorporating such a feature into AVA is a wonderful one.''}
By aligning with norms of human honesty, abstention laid the groundwork for richer anthropomorphic interpretations.
We observed two distinct modes of anthropomorphizing. First, descriptive anthropomorphizing emerged as participants characterized AVA through personality traits inferred from its behavior.

For example, P2 depicted AVA as a reserved yet efficient colleague:
\begin{quote}\itshape
``If I think of AVA talking as a human being, AVA is a bit reserved, a person of few words. AVA doesn't want to force words into your mouth; AVA wants you to think for yourself. She's unique and also very efficient.''
\end{quote}

P2 further characterized AVA's abstention as deliberate encouragement for independent thinking, contrasting this with the verbosity of generalist LLMs: \textit{``For ourselves, we think and we get to the answer. But for ChatGPT, GPT wants to give you the first word. GPT thinks for you.''}  Second, aspirational anthropomorphizing reflected participants' desires for enhanced interaction qualities. P13 wished for AVA to feel like a closer companion: \textit{``Humanize it, make it human, make it as my fellow, make it just my brother\ldots{} And make it learn to reflect the individual person who is using it.''} He suggested softening abstentions with service-oriented phrasing, replacing \textit{``I don't know about this''} with \textit{``I am still learning about this. Is there something else I can help with for now?''}, to project a more helpful, service-oriented persona that positions the AI as a cooperative partner rather than a simple tool.

\noindent
Thus, anthropomorphization functioned as a sensemaking strategy, rendering system constraints legible and workable (predictability, calibrated reliance, honest), while simultaneously providing a vocabulary for articulating unmet social needs (warmth, rapport).

\subsubsection{Balancing Honesty with Usability}

While the majority of participants appreciated AVA's principled abstention, some identified it as a workflow bottleneck. A non-answer, however well-justified, still left their information need unmet. In response, these participants suggested alternative approaches.

For example, P7 described how AVA could leverage its corpus knowledge for structured guidance, proposing that when the system finds related but imperfect matches, it should use those results to help users formulate more targeted queries:
\begin{quote}\itshape
``Maybe, for example, in this case that I wasn't able to find the information that I was looking for, but then AVA sent me this great information from another region of the world. If the tool can give you not only the suggestion, but refine the prompt with you and say: `I couldn’t find information for your specific region, but you could try refining your search. Can you provide more details, such as a year, a specific mineral, a community, or a particular government agency? ''
\end{quote}

\noindent
This desire for collaboration also extended to the interaction's conversational style. P9 focused on the emotional tone of the interaction, arguing that the bluntness of an ``I don't know'' response felt ``rude''. Instead, he proposed a more engaging, suggestive approach to maintain the conversational flow: \textit{``It could come up with, `do you mean?’... Or you can look out for this from this... that kind of trying to engage you a bit... `I don't know’ sounds so rude.''} This pattern reframes the interaction as a helpful clarification rather than a conversational dead end.

\noindent
Finally, when query refinement was insufficient, participants wanted AVA to act as a knowledgeable guide to the broader information ecosystem. Participants proposed that AVA should guide them to external resources, such as other LLMs or web results. 
\noindent
These suggestions reveals participant's vision for a more collaborative AI that pairs honest refusal with external handoffs to reduce workflow friction. While we present these findings here, we analyze their deeper implications for trust and multi-tool ecosystems in the Discussion.

\subsection{Trust Calibration in Evidence Practices (RQ2)}
Trust in AVA was calibrated rather than assumed. We adopt Gambetta’s definition of trust~\cite{gambetta2000trust} as the belief that another agent’s actions are sufficiently beneficial, or at least not harmful, to justify cooperation. Participants did not take outputs as inherently reliable; rather, they assessed trust through two intertwined layers. At the feature level, page-anchored citation verification and reasoned abstention functioned as reliability cues, indicating when claims were grounded and when the system should refrain. At the dataset level, confidence derived from AVA’s curated corpus of World Bank Reports. The subsections that follow examine how these feature- and dataset-level mechanisms informed participants’ decisions to rely on AVA in evidence work.

\subsubsection{Feature-Level Trust}

Participants located trust in two key interface mechanisms: page-anchored citation verification and reasoned abstention.
\noindent
First, citation verification was highly valued, with 17 of 20 participants rating it 5/5 in usefulness and the remainder rating it 4/5. Participants reported selective verification rather than exhaustive checking, typically when claims were surprising, counterintuitive, or numerical. Across interviews, participants reported no mismatches between claims and sources. Qualitatively, participants contrasted AVA's page-level, context-anchored citations with the opaque ``link dumps'' of general-purpose systems. As P17 noted:
\begin{quote}\itshape
``This commitment to bring on the sources and say it comes from there. If we compare with other AI engines, they give you everything which has been used by the model and then a couple of links at the bottom, and good luck for you to link or address it.''
\end{quote}
P20 emphasized that seeing source context was central to perceived reliability, as it is \textit{``very helpful to see where the data comes from and also the context in which the data was presented, which reassures you the model does not hallucinate.''}

Second, reasoned abstention operated as a complementary trust mechanism by preventing overclaiming, particularly in high-stakes contexts. As noted in Section 6.2.1 , participants (e.g., P7) contrasted AVA’s cautious non-response with the overconfident behaviour of general-purpose systems. This honest abstention helped participants calibrate their reliance on the system, distinguishing AVA from tools that might overstate confidence.

\subsubsection{Dataset-Level Trust}

Source credibility was enhanced by corpus provenance. Participants emphasized that AVA's foundation in a curated library of official World Bank Reports provided inherent trustworthiness, as many had previously relied on this repository in their professional work. P5, a researcher, noted his long-standing relationship with World Bank Reports, mentioning he has been using them since his undergraduate studies. This prior familiarity with institutional sources reduced concerns about data quality independent of AVA's interactive features. P1, a lecturer, articulated this trust in institutional authority: \textit{``It's coming from World Bank or World Health Organization or UNDP. I know that this [is] authentic. So I prefer even getting my information from this site than every other out there''}. Similarly, P6, a policy analyst, noted: \textit{``Because I know World Bank is the best institution for reports and research... if I'm putting [it], I don't have to think that I haven't done enough research because I know those who put it there are the originators''}.

P12 positioned AVA as competitive with major AI platforms specifically due to its curated approach:  

\begin{quote}
``I feel AVA has the capability of being on the same page as ChatGPT...  
AVA has an advantage because its information is not just picked up internet-wise, [it's] actually concrete information.'' 
\end{quote}

\noindent
Overall, participants noted feature-level mechanisms and dataset provenance were mutually reinforcing. Citation affordances are only as credible as the materials they point to; conversely, a reputable corpus attains practical value when claims are traceable to page-level evidence. 

\subsection{Contested Approaches to Disclosure (RQ4)}
Participants articulated sophisticated yet conflicting perspectives on whether AVA's assistance should be disclosed in professional documents, often drawing comparisons with other AI tools to clarify their positions. Such comparative reasoning is consistent with prior work in HCI and the social sciences, where individuals use analogies to familiar systems to make sense of novel or ambiguous technologies and to articulate emerging norms and boundaries~\cite{schwarz2018power,schwarz2018nanotechnology, ratzan2000making}
These accounts reflected an ongoing negotiation of emerging professional norms rather than adherence to established ethical guidelines. While few participants ( P3 and P6) explicitly argued that disclosure was necessary, others expressed more ambivalent or opposing views. Across this diversity of opinion, four distinct orientations emerged, which we detail below.

\subsubsection{Disclosure as Strategic Advantage}  
Some participants framed disclosure as an opportunity to enhance transparency and credibility. P12 argued that institutional provenance could strengthen the legitimacy of outputs: \textit{``Oh, yes. I feel it should be disclosed. It builds transparency\ldots{} If I go to a meeting with my research findings and I say the AI is based on World Bank Reports and everybody knows the World Bank is a big entity and all the information is correct, it gives me some kind of positive impression.''}  
P1 similarly noted that disclosure could promote broader AVA adoption, reframing it as an advantage rather than a liability.

\subsubsection{Positioning AI Use as Ordinary Practice}  
Others challenged disclosure requirements altogether, framing AI as a routine professional aid that do not require explicit acknowledgement. P13 drew an analogy to secretarial assistance: \textit{``I have a secretary and my secretary is helping to type a letter. So should I tell the person I sent the letter this has been written by my secretary?''} He described AI as a seamless co-creator: \textit{``I consider AI as a tool, as a part of me. We have created it together.''}

\subsubsection{Contextual and Domain-Specific Norms}  
Several participants emphasized situational nuance. P9 distinguished between academic contexts, where disclosure was expected, and applied development settings, where effectiveness mattered more: \textit{``If I can lift the whole document and implement it here and it works, maybe change one or two things. For me, I don’t see it as anything. Disclosure is important for academic writing though.''} Others, such as P14 differentiated computational from intellectual AI support, suggesting that while data processing might not require disclosure, creative writing did, due to plagiarism concerns.

\subsubsection{Reframing Disclosure as Professional Integrity}  
Finally, some participants shifted the focus from rules to professional integrity. P11 critiqued denial of AI use as disingenuous: \textit{``Anyone saying "I don’t use AI for anything", then I think it’s just a lie. The world has adapted to it already.''} He argued that the true ethical imperative was to build competence: \textit{``You need a lot of work to actually know how to use AI tools properly.''} P20, who viewed AI as a thought partner rather than a copy-paste substitute, concurred, noting that while AI carries risks, \textit{``the human reviewer has already validated and removed those risks,''} making disclosure unnecessary once content had been vetted.

In sum, these perspectives illustrate professionals actively constructing contextual ethics around AI disclosure, balancing strategic advantage, seamless integration into practice, domain-specific norms, and evolving standards of honesty in an AI-ubiquitous landscape.

\subsection{Triangulation of Quantitative and Qualitative Findings}

Our mixed-methods strands converge on three points. First, quantitative analyses (Tables~\ref{tab:summary_Of_Policy_Tasks_Reflected_in_Queries}, \ref{tab:Summary_statistics_of_characteristics_of_registered_users_and_users_with}, \ref{tab:user_agreement}, \ref{tab:regression_results}, \ref{tab:did_results}) show broad engagement patterns and strong associations with self-reported efficiency (80.5\%) and work quality (76.3\%) (See Table \ref{tab:user_agreement}). Interviews explain the mechanisms underlying these high ratings, namely page-anchored citation verification as an evidence shortcut. This feature was rated 5/5 on usefulness scale by 17 of 20 participants and calibrated reliance via reasoned abstention. Second, dose--response patterns among returning users showing time savings ($b=1.838$, $p<.10$; $b=3.272$, $p<.10$) complement the \textit{ITT} null results ($b=-0.568$, $p>.05$; $b=-0.777$, $p>.05$); interviews clarify when repeated use was associated with efficiency gains through multi-tool orchestration and verification routines that develop over time. Third, quantitative indicators of high trust (82.2\% found AVA trustworthy) (See Table \ref{tab:user_agreement} and recommendation rates (89.0\%) (See Table \ref{tab:user_agreement}) align with interview accounts of provenance-first interaction grounded in World Bank institutional credibility.

We also observe productive divergences. Some participants experienced abstention as a workflow bottleneck despite overall high satisfaction (72\%); where available, logs and interview reports contextualize exposure to abstention events, motivating design implications (guided reformulation, external handoffs). Finally, English-dominant usage in logs (83.8\% of interactions) is consistent with interviews citing workplace norms rather than quality concerns. We synthesize these meta-inferences and design implications in the Discussion.

\section{Discussion}

We draw on data from a large-scale, in-the-wild global deployment of AVA used by more than 2,000 policy professionals across diverse organisations in 116 countries for 5 months, supplemented by 20 in-depth qualitative interviews (See Figure \ref{fig:multi-institional}). In this paper, we use AVA as a case study to examine how established techniques such as corpus curation~\cite{oche2025systematic}, reasoned abstention~\cite{wester2024denials,kim2024uncertainty}, and page-level verification~\cite{huang2024learning,xia2025ground}, function when implemented in a real-world, high-stakes setting, rather than to claim novelty of the underlying architecture~\cite{wen2025know}. The contribution we offer is therefore empirical and conceptual: we situate AVA’s value in insights from the deployment about how policy professionals engaged with an evidence-bounded generative system in their day-to-day work and the professional practices that formed around it, spanning appropriation(RQ1)~\cite{dourish2003appropriation} , trust calibration (RQ2)~\cite{lee2004trust}, long-term use and perceived efficiency gains (RQ3~\cite{orlikowski1995evolving}, and norms of disclosure and accountability (RQ4)~\cite{rezaei2024ai,formosa2025can}.

\textbf{The Necessity of Agentic Complexity}: Before detailing specific design lessons, we address the architectural necessity of the multi-agent retrieval system (Stage 2). While we observe that user trust is grounded primarily in the curated corpus (Stage 1) and the verification pipeline (Stage 3), we posit that the intermediate agentic overhead (Stage 2) is operationally essential in the context of policy and development. Policy analysis queries frequently require synthesizing dispersed evidence across documents and languages, conditions under which simpler, single-agent RAG pipelines fail to maintain coverage and precision~\cite{dalglish2020document}. The multi-agent structure enables the system to plan, navigate document hierarchy, and aggregate diverse sources more reliably, thereby providing the necessary evidence pool for Stage 3 verification to ensure evidentiary correctness.

In this section, we synthesize these insights into three transferable design lessons for building specialised AI systems in other high-stakes domains, such as healthcare, legal practice, and governance. We then introduce a fourth, forward-looking design recommendation: an interoperable AI ecosystem that supports coordination across specialised and general-purpose AI systems.

\subsection {Lesson 1: Designing Specialized AI Systems for High-Stakes Knowledge Work Requires an End-to-End Trust Pipeline}

Our empirical findings indicate that AVA’s successful adoption was rooted in two principles that address the specific epistemic needs of policy and research professionals: Dataset-Level Trust (grounded in the curated World Bank corpus) and Feature-Level Trust (encompassing reasoned abstention and clickable citations). Specifically, the observed Feature-Level Trust decomposes into mechanisms residing at the Model Layer (reasoned abstention) and the Interface Layer (citation verification). In the discussion below, we synthesize the idea of a trust-pipeline that spans the entire system architecture.

\begin{enumerate}
\item 
\textbf{Dataset-Level Trust (Curation):}
The system established dataset-level trust by grounding its responses in a curated library of World Bank Reports, a corpus with which participants were already familiar and whose institutional provenance they respected. Trust was therefore inherited from established professional practice.

\item 
\textbf{Model-Level Trust (Abstention):}
In AVA, the model behavior shapes how inherited trust from the curated dataset is preserved. AVA’s reasoned abstention  mechanism, implemented through a fixed evidence-verification threshold, constrained the system to answer only when sufficient supporting evidence was available, otherwise abstain. Most participants interpreted “I don’t know” responses not as system failures, but as boundaries of the system and as preferable to the speculative or fabricated answers they commonly encountered in generic AI systems.  This evidence-sensitive response strategy functioned as a critical safeguard in a context where speculative or fabricated answers  could have significant downstream consequences.

\item 
\textbf{Interface-Level Trust (Verification):}  
Finally, trust is enacted at the interface, where verification features allow users to inspect AVA's outputs against the trusted corpus. Page-level, clickable citations enabled participants to trace specific claims back to their sources. Seventeen of twenty interview participants rated this feature at the highest level (5/5), even though most reported using citation checking selectively, primarily to calibrate trust when an output was surprising, counterintuitive, or numerically specific in high-stakes analytical work.
\\
While this pipeline establishes structural trust mechanisms, user experience requires further nuance. Our deployment revealed that rigorous grounding must balance social expectations and residual error. We detail these two interactional challenges below:

\textbf{Tensions between epistemic humility and relational humility:} Adapting the construct of \textbf{relational humility }~\footnote{In interpersonal psychology~\cite{davis2013humility}, relational humility is defined as an observer’s judgment that a person within a relationship demonstrates three qualities: (1) an accurate view of the self (neither inflated nor diminished), (2) an other-oriented stance that prioritizes the welfare of interaction partners, and (3) interpersonal behaviors marked by a lack of superiority or the regulation of ego-focused emotions (e.g., modesty).} from interpersonal psychology to AI, we conceptualize relational humility as a user-attributed interactional property of the AI system that emerges when it: (1) communicates its limits; (2) remains oriented to the user’s goals, for example by explaining why an answer cannot be verified and suggesting constructive next steps; and (3) adopts collaborative, non-superior language. While epistemic humility concerns acknowledging limits in knowledge and deciding when to abstain, relational humility concerns the interactional stance through which those limits are communicated and managed to preserve collaboration and trust. As noted in Section~\ref{sec:abstention_findings}, some participants interpreted abstentions as `rude' or a violation of partnership, expressing a desire for a `brother-like' persona that felt more humanised. 
This suggests that, in professional settings, a bare refusal (e.g., `I don't know') risks violating the Cooperative Principle of Conversation. To mitigate this, refusals should function as ``service-oriented'' pivot
(e.g., `I cannot verify X, but I can help you explore Y'). Consequently, the design lesson is to pair rigorous abstention with relational humility as part of the end-to-end trust pipeline. 
Future systems should match strict evidence-based refusal with a relationally humble persona that performs \textbf{social repair}, so that refusals reduce friction and preserve users' trust even when the AI reaches its limits.

\textbf{The inevitability of residual error:} Even with a robust trust pipeline, designers must acknowledge that specialized systems will still occasionally surface incorrect or mismatched evidence~\cite{10.1145/3626314, zhou2023synthetic}.  If the system is highly trusted, the pipeline can paradoxically lead to \textit{automation bias}~\cite{10.1145/3449287, lyell2017automation}, where users stop verifying claims because the system is usually right. Therefore, even when verification is distributed across the trust pipeline, designers should assume residual error will persist and incorporate interface cues, such as uncertainty signals, contrastive evidence previews, source comparisons, in-context highlighting, or interactive hover previews that make identifying inconsistencies low-friction \cite{braun2024humans, sun2025explaining, zhang2021manage}. 
Future researchers could explore more domain-specific ways to highlight such residual errors.

\end{enumerate}

\subsection{Lesson 2: Balancing the Source Quality and Coverage Tradeoff}
A core open question in building AI-powered knowledge systems is how to trade off coverage (the breadth of topics a system can address) against source quality~\cite{digiacomo2025guide, oche2025systematic}. General-purpose models such as ChatGPT illustrate the risks of maximising coverage over source quality, as they aggregate content from the open Internet~\cite{oche2025systematic}. OpenAI's recent policy guidelines, released in October 2025, state that ChatGPT will not provide recommendations in domains like law and medicine~\cite{openai_usage_policies}; however, it still provides answers while appending lightweight disclaimers~\cite{businessinsider_chatgpt_health_2025}. In these settings, the issue is not the amount of evidence; there is abundant text supporting almost any claim on the Internet, but rather the quality and verifiability of that evidence, which then requires substantial, often manual, verification effort. This motivates an alternative design choice for building specialized AI systems for high-stakes knowledge work: constrain coverage to high-quality, auditable sources, and expand only when justified by clear signals of unmet user needs or corpus–task mismatch.

In AVA, we instantiate this quality–coverage trade-off by fixing the evidence-verification threshold and treating abstention as the primary observable signal of corpus adequacy. In the initial deployment (See figure~\ref{fig:5_day_total_queryvolume}), AVA operated over a narrowly scoped corpus of roughly 50 flagship World Bank Reports. This maximized source quality but yielded high abstention rates (40–70\%). Interviews indicated that repeated abstentions prompted some users to switch to alternative tools, reflecting a clear corpus–task mismatch. Guided by the distribution of unanswered real-world queries, we expanded AVA’s corpus, within the same vetted institutional boundaries, to more than 4,000 World Bank Reports, leaving the verification threshold unchanged. Under this broader yet still curated corpus, abstention fell to approximately 10\% (See figure~\ref{fig:5_day_total_queryvolume}). System logs show that AVA addressed the majority of queries in this phase, and participants interpreted the remaining abstentions positively, contrasting AVA’s explicit “I do not know” with other AI tools that “make things up.” Once coverage was sufficient, abstention shifted from being perceived as a limitation to functioning as evidence of caution.

This leads to our core design principle: abstention  should reflect epistemic humility rather than systematic failure. Because high factual reliability is unlikely to be achievable when systems rely on unbounded, weakly verifiable sources such as the open Web~\cite{businessinsider_chatgpt_health_2025, lin2022truthfulqa}, we constrain AVA to a curated corpus and enforce a fixed evidence-verification threshold. In doing so, we accept that the system will sometimes abstain as a necessary trade-off for reliability. To disambiguate abstention signals, we distinguish abstention (system property) from corpus–task alignment (coverage adequacy). Building on this distinction, we recommend that future systems monitor three complementary signals in tandem with the abstention rate: corpus–task alignment , evidence verification threshold and user behaviour (including reformulations, abandonment, and patterns of re-engagement). Interpreted together, these signals can help developers decide when to expand a curated corpus or tighten evidence-verification thresholds, supporting a dynamic, evidence-grounded approach to managing the quality–coverage trade-off while avoiding the pitfalls of open-web sourcing.

\subsection{Lesson 3:  Design for Verification, Not Just Disclosure}

Participants’ approaches to disclosure (RQ4) revealed a hierarchy of needs: while disclosing AI use was seen as contextual, ensuring the accuracy of AI outputs was viewed as non-negotiable. This distinction reframes disclosure not as an end in itself, but as secondary to verification in high-stakes professional practice. It reveals a critical divergence between the institutional emphasis on disclosure (transparency of method)~\cite{hoque2024hallmark,Sun2024MetaWriter} and the professional necessity of verification (accuracy of output). This has direct implications for the design of specialised AI systems. It is insufficient to merely provide the capacity to verify; designers must ensure that verification mechanisms streamline the process rather than creating friction. If verification imposes a high cognitive cost, users are likely to succumb to automation bias and bypass verification altogether~\cite{10.1145/3449287, lyell2017automation}. AVA addresses this through features like page-level verifiable citations and reasoned abstention, which function as tools for seamless answer verification. By enabling users to inspect underlying evidence without breaking their workflow, these mechanisms align with our participants’ preference for tools that make system fallibility visible and actionable.

\subsection{Lesson 4: Embracing the Future: From a Single Tool to a Collaborative AI Ecosystem}
\label{sec:discussion_ecosystem}

Our findings suggest that AVA’s perceived value derived predominantly from providing reliable, citable evidence. This success overshadowed challenges such as  limited feature discoverability. AVA’s specialization precluded open-ended tasks like brainstorming. leading users to routinely turn to general-purpose AI tools. We view this switching not as failure but as evidence of an emergent collaborative AI ecosystem. This model mirrors the long-standing practice of knowledge workers who employ a suite of distinct software applications, such as spreadsheets, word processors, and presentation software, to accomplish a complex task~\cite{jahanlou2023task}. Consequently,  design should foster seamless interoperability. \cite{Jackson2014PolicyKnot}.

\subsubsection{Design Implications: The Intelligent Handoff as a Cornerstone of Interoperability}
A central challenge of this ecosystem model is managing the transition between tools without degrading user trust. If a trusted specialist AI refers a user to a generalist AI that subsequently provides incorrect information, the specialist's reputation may be harmed. To mitigate this risk, we propose the "Intelligent Handoff" as a design pattern for responsible interoperability. A successful handoff should not simply be a link, but a structured interaction that performs three key functions:
\begin{enumerate}
    \item \textbf{Communicate Boundaries and Manage Expectations:} It should clearly state \textit{why} it is abstaining (e.g., "I cannot answer this from my verified library...") and frame the recommendation by task type (e.g., "...for creative brainstorming, a general-purpose AI may offer starting points").
    \item \textbf{Preserve User Agency:} Rather than prescribing a single product, the handoff should recommend a \textit{class} of tool, empowering the user to select the service that best fits their needs and context (e.g., "...you could use services like ChatGPT, Gemini, or Claude.").
    \item \textbf{Add Value Through Prompt Generation:} To increase the likelihood of a successful outcome, the specialist AI can use its domain knowledge to provide a re-engineered, higher-quality prompt for the user to copy. This adds immediate value, reduces the risk of hallucination, and reinforces the specialist AI’s role as a helpful, expert collaborator.
\end{enumerate}

\textbf{Generalizability to High-Stakes Knowledge Work}: 
Our design principles are not specific to international development and policy context and applies to other high‑stakes domains such as law and medicine, where evidentiary risk and professional accountability are central. We outline concrete implications for these domains in Appendix \ref{app:transferability}.

In sum, our work suggests that the future of AI for knowledge work lies in creating specialized tools that are powerful in their own right and designed to be excellent, responsible collaborators within a broader digital ecosystem. Ultimately, our work extends the concept of Humble AI~\cite{knowles2023humble}. While prior research~\cite{knowles2023humble} emphasizes systems communicating their internal limits, our findings indicate that, for specialized tools, humility must also be ecosystem-aware. A responsible AI collaborator does not stop at its own knowledge boundary; it offers users intelligent, trust-preserving pathways to other tools. This shift from introspective humility to collaborative humility is critical for designing AI systems that integrate seamlessly and responsibly into the complex, multi-tool reality of professional knowledge work.

\section{Limitations}


Our study has several limitations that should be considered when interpreting these findings. First, the five-month deployment limits conclusions about long-term usage patterns and downstream impacts. Practices around trust calibration, system appropriation, and multi-tool orchestration may evolve as users gain familiarity with AVA and as organizational norms around AI assistance stabilize. Second, our multilingual evaluation was constrained by participants' English-dominant workplace practices. While two participants provided positive feedback on Spanish and French outputs, the small non-English sample (83.8\% of interactions were English) limits conclusions about AVA's performance across its 60+ supported languages or in predominantly non-English research contexts. Future research would focus on evaluating these languages with users. A third limitation of the study is the potential self-selection bias among participants who completed the endline survey and interviews. Prior work notes that such self-selection is a common characteristic of in-the-wild, ecologically valid deployments~\cite{andrade2018internal,ram2017questionable}.
To help mitigate this issue, we sent multiple reminder emails and conducted interviews with volunteers representing diverse usage patterns and professional backgrounds. Still, voluntary participation may nonetheless introduce bias toward more engaged users. At the same time, we complement findings based on surveys and interviews with user log data, which provides more objective behavioral indicators of engagement. Future research might address this limitation through targeted recruitment strategies to capture less active users.

\section{Conclusion}
This paper reports a five-month, in-the-wild deployment of AVA, a curated, citation-grounded multi-agent AI assistant for policy and development professionals. Using mixed methods, we show how page-level verifiability, reasoned abstention, and institutional provenance reshape evidence practices, enabling AVA to function as an evidence engine that improves efficiency while reducing reliance on open-web LLMs for high-stakes analysis. We distill three design lessons: trust must be engineered as an end-to-end pipeline spanning corpus curation, model behavior, and interface-level verification; the quality–coverage trade-off must be deliberately governed, with abstention treated as a signal of reliability rather than failure; and system design should prioritize low-friction verification of AI-generated outputs rather than disclosure alone. Ultimately, we advocate for ecosystem-aware humility in generative AI, where systems clearly bound their claims and enable responsible handoffs when tasks exceed their scope.


\begin{acks}
We thank the anonymous reviewers for their constructive feedback. We also thank Ali Moezzi for his contributions to AVA. We extend our gratitude to Dr. Bill Thies and Professor Ge "Tiffany" Wang for their insightful discussions during the revision process. Finally, we thank Professor Sir Nigel Shadbolt and Professor Max Van Kleek for their support. The findings, interpretations, and conclusions expressed in this paper are entirely those of the authors. They do not necessarily represent the views of the International Bank for Reconstruction and Development/World Bank and its affiliated organizations, or those of the Executive Directors of the World Bank or the governments they represent. © 2026 International Bank for Reconstruction and Development/International Development Association or The World Bank This work is provided under a Creative Commons 4.0 Attribution International License, with the following mandatory and binding addition:
Any and all disputes arising under this License that cannot be settled amicably shall be submitted to mediation in accordance with the WIPO Mediation Rules in effect at the time the work was published. If the request for mediation is not resolved within forty-five (45) days of the request, either You or the Licensor may, pursuant to a notice of arbitration communicated by reasonable means to the other party refer the dispute to final and binding arbitration to be conducted in accordance with UNCITRAL Arbitration Rules as then in force. The arbitral tribunal shall consist of a sole arbitrator and the language of the proceedings shall be English unless otherwise agreed. The place of arbitration shall be where the Licensor has its headquarters. The arbitral proceedings shall be conducted remotely (e.g., via telephone conference or written submissions) whenever practicable, or held at the World Bank headquarters in Washington DC.

\end{acks}

\bibliographystyle{ACM-Reference-Format}
\bibliography{sample-base}

\clearpage
\onecolumn
\appendix

\newpage

\appendix
\section{Detailed Descriptions of AVA Interface Components}
\label{app:walkthrough}

Figure~\ref{fig:ava-walkthrough} presents an annotated view of AVA’s interface. 
Below we provide detailed descriptions of each labeled component (A–H).

\textbf{A. Query Input Box.}
The user enters a natural-language research question. This initiates the retrieval and grounding pipeline.

\textbf{B. Retrieval Process Trace.}
A dynamic trace (e.g., “Thinking through 186 sections and 10 images”) that surfaces the system’s retrieval operations, indicating how AVA scans, ranks, and selects evidence from the curated corpus to ground its response.

\textbf{C. Inline Verifiable Citations.}
Clickable, page-level citation markers embedded directly within the generated answer. These act as entry points into the verification loop.

\textbf{D. Evidence Preview Card.}
A pop-over that appears when the user selects a citation. It displays the document title, page number, and a brief snippet of surrounding text to allow quick, in-context inspection without leaving the chat interface.

\textbf{E. Source Document Viewer.}
A side-by-side PDF viewer that opens when the user clicks the document header. The viewer automatically scrolls to the cited page and highlights the referenced text, supporting low-friction verification.

\textbf{F. Save-as-Note Functionality.} A feature enabling users to save the generated response and the verified citations into a dedicated notes panel for later reference.

\textbf{G. Copy-Paste Functionality.}
A control that allows users to copy the generated text (including citations) into external tools such as Word, Google Docs, or email workflows.

\textbf{H. Tagging.}
Tagging features that allow users to assign labels to saved notes or research sessions, supporting organisation and later retrieval.

\section{Consolidated Usage Patterns, Longitudinal Dynamics, and User Insights} 
\label{sec:appendix_usage_patterns}

Appendix B consolidates empirical usage patterns, longitudinal dynamics, and user-reported friction points that relate to behavioural findings. 
It brings together usage statistics, temporal shifts, and interview-reported challenges that are otherwise distributed across Sections 5 and 6, providing a single reference point for these interactional patterns. Because RQ4 concerns sociocultural disclosure norms rather than system usage behaviour, those findings remain in the main Results (Section 6.4).

\subsection{Indicative Usage Patterns}
Our mixed-methods analysis revealed three dominant patterns of engagement:
\begin{itemize}
    \item \textbf{Task \& Theme Dominance:} Usage was heavily skewed toward \textit{Diagnostic} tasks (69.0\% of queries), such as understanding problems or data, rather than Design (22.0\%) or Evaluation (9.0\%). Thematically, users focused on broad, cross-cutting policy areas like \textit{Human Capital} (33.5\%) and \textit{Macroeconomics} (20.7\%) rather than niche fact retrieval.
    \item \textbf{Multi-Tool Ecosystem:} Users exhibited a distinct ``multi-tool'' workflow, utilizing generalist LLMs (e.g., ChatGPT) for ideation and scaffolding, while reserving AVA specifically as an ``evidence engine'' for ``real information'' and citations.
    \item \textbf{Language Norms:} Despite supporting 60+ languages, 83.8\% of interactions occurred in English. Interviews revealed this was driven by workplace norms in international development rather than a lack of capability.
\end{itemize}

\subsection{Longitudinal Shifts in System Behavior}
We observed two major shifts over the five-month deployment:
\begin{itemize}
    \item \textbf{Impact of Corpus Expansion on Abstention:} System behavior changed significantly over time. Initially, with a small corpus (50 reports), the system abstained on 40--70\% of queries. After expanding to 4,000+ reports, abstention dropped to below 10\%. This shifted the user experience of abstention from a ``symptom of mismatch'' to a ``positive signal'' of caution.
    \item \textbf{Efficiency Gains for Returning Users:} The study found a dose-response relationship where returning users (multiple sessions) reported significant time savings (2.4--3.9 hours/week) compared to single-session users, suggesting that efficiency gains and verification routines develop with sustained use.
\end{itemize}

\subsection{User Frustrations and Challenges}
While satisfaction was high, specific friction points emerged in qualitative interviews:
\begin{itemize}
   \item \textbf{The ``Dead-End'' Abstention:} 
While many participants appreciated abstention as an honest boundary, others experienced frequent “I don’t know” responses as a bottleneck that stalled their workflow. In these moments, users reported turning to other tools to maintain momentum.
    \item \textbf{Tone \& Usability:} Specific users described the bluntness of the refusal as ``rude''. Others expressed frustration that the system stopped completely rather than offering partial help, stating, ``I don't want the choice being no answer''
\end{itemize}

\subsection{Suggested Improvements}
Participants proposed three specific enhancements to better support high-stakes workflows:
\begin{itemize}
    \item \textbf{Intelligent Handoffs:} To address the frustration of dead ends, participants suggested the system should act as a ``knowledgeable guide'' that points them to external resources (e.g., ``try ChatGPT for this'') when it cannot answer from the verified corpus.
    \item \textbf{Collaborative Query Refinement:} Users requested that the system help them refine their prompts or suggest related regions/years when exact matches are missing, rather than just abstaining.
    \item \textbf{Exportable Visuals:} Users praised the generated charts and tables but requested better functionality to export these visuals with built-in attribution for reports.
\end{itemize}

\section{Supported Languages}
\label{app:langauge}
AVA currently supports input and output in the following 60+ languages: Albanian, Amharic, Arabic, Armenian, Bengali, Bosnian, Bulgarian, Burmese, Catalan, Chinese, Croatian, Czech, Danish, Dutch, English, Estonian, Finnish, French, Georgian, German, Greek, Gujarati, Hindi, Hungarian, Icelandic, Indonesian, Italian, Japanese, Kannada, Kazakh, Korean, Latvian, Lithuanian, Macedonian, Malay, Malayalam, Marathi, Mongolian, Norwegian, Persian, Polish, Portuguese, Punjabi, Romanian, Russian, Serbian, Slovak, Slovenian, Somali, Spanish, Swahili, Swedish, Tagalog, Tamil, Telugu, Thai, Turkish, Ukrainian, Urdu, and Vietnamese.

\section{Extended Methodology and Measures}
\label{app:methodology_extended}

\subsection{Recruitment, Ethics, and Study Timeline}
Recruitment and Ethics: Participants were recruited through public-facing newsletters and targeted emails from the host MDB. Participation had no bearing on employment status, course enrollment, or standing with the institution. All activities were conducted on participants' own devices to mirror heterogeneous real-world bandwidth conditions.

Detailed Timeline: \begin{itemize}
\item \textbf{Baseline Data Collection:} April to May 12. 
\item \textbf{Pilot Launch:} May 12 (small cohort). 
\item \textbf{Full Rollout:} May 26 (expanded to all randomly selected treatment registrants). 
\item \textbf{System Updates:} During deployment, we shipped corpus expansion to >4,000 documents (July 9), graph-generation features (July 17), and improved handling of non-response queries (July 23). 
\item \textbf{Endline Survey:} Administered July 21. 
\item \textbf{Public Launch:} August 4. 
\end{itemize}

\subsection{Survey Instruments and Variable Definitions}

\begin{enumerate}
  \item \textbf{AI Use for Work (Baseline \& Endline).}
  AI experience was measured via a six-point frequency scale: “Multiple times every day,” “About once or twice on most days,” “A few times throughout the week,” “Rarely,” “Not at all,” and “Never use AI for work.”

  \item \textbf{AI-Driven Productivity Gains (Baseline \& Endline).}
  Respondents estimated hours saved in the prior week (writing, research, analysis, formatting, summarizing) with categories: 0, $<$1, 1–3, 4–6, 7–10, $>$10 hours. We compute lower and upper bounds by mapping ranges to their endpoints (e.g., 1–3 $\rightarrow$ 1 and 3; 4–6 $\rightarrow$ 4 and 6; 7–10 $\rightarrow$ 7 and 10).

  \item \textbf{Perceptions of AVA (Endline).}
  Among users granted access, we assessed: efficient access to insights from World Bank flagship documents (“AVA helps me understand key insights from World Bank flagship documents more efficiently.”), trustworthiness/accuracy (“The synthesized insights provided by AVA are accurate and trustworthy.”), work-quality impact (“Using AVA has improved the quality of my work or research.”), and recommendation intent (“I would recommend AVA to others working in development policy or research.”).

  \item \textbf{Perceptions of AVA (In-App).}
  Periodic session pop-ups captured the same perceptions (e.g., trustworthiness, likelihood to recommend). We report perception results from both endline and in-app sources for a comprehensive view.
\end{enumerate}

\subsection{Detailed NLP Preprocessing and Query Classification}

\paragraph{Preprocessing Pipeline}
\begin{itemize}
    \item \textbf{Translation \& Normalization:} Language detection was applied to every query; non-English text was automatically translated to English. Text was lowercased, and punctuation, digits, and extraneous whitespace were removed.
    \item \textbf{Tokenization:} Common English stopwords and very short tokens were filtered out. Remaining tokens were lemmatized to their base forms.
    \item \textbf{Sessionization:} Queries were grouped into sessions using a strict one-hour inactivity threshold to capture immediate task context.
\end{itemize}

\paragraph{Classification Logic}
\begin{itemize}
    \item \textbf{Taxonomy Matching:} Classification was rule-based using three curated taxonomies (Policy Themes, Query Types, Intended Use). We used weighted keyword matching where multi-word phrases received higher priority than single-word matches.
    \item \textbf{Conversational Filtering:} A separate conversational keyword list was checked first to identify and segregate chit-chat or phatic interactions.
    \item \textbf{Imputation:} A TF-IDF step surfaced representative terms. Finally, uncategorized queries were forward-filled based on neighboring categorized queries within the same session to leverage conversational context.
\end{itemize}

\subsection{Positionality Statement}
We acknowledge that our backgrounds shaped both AVA-AI's design and evaluation. Our team spans human--computer interaction researcher (4--8 years), development policy practitioner  (8--15+ years across education, climate, and social protection), and machine learning engineer (3--6 years in retrieval systems). Several authors have first-hand policy experience and working within public-sector constraints; this vantage influenced our emphasis on citation-first interfaces, claim-level verification with coverage thresholds, and workflow integration aligned with policy research practice.

We recognize this may bias us toward formal, documentable sources and structured evidence over other forms of knowledge (e.g., gray literature, oral testimony). To counterbalance, we evaluated AVA with users in diverse roles across multiple countries, reported failure cases and abstention triggers alongside successes. Our motivation stems from direct experience with information overload in development research, synthesizing large literature while maintaining source traceability under time pressure. We aim to contribute design patterns that augment expert judgment in knowledge-intensive domains.

\section{Transferability of Design Principles to Medical and Legal Workflows}
\label{app:transferability}
Recent changes by Open AI restrict the use of general-purpose models for tailored legal and medical advice, formally acknowledging the evidentiary and safety risks of using LLMs in these domains~\cite{openai_usage_policies}. Yet, in practice, ChatGPT still answer legal and medical questions while appending only lightweight disclaimers~\cite{businessinsider_chatgpt_health_2025}, placing the burden of verification entirely on users. This design pattern risks being counterproductive: it normalises confident answers in domains where errors can be catastrophic, while offering minimal support for systematic checking. Prior work on AI-assisted decision making shows that when verification is costly or poorly supported, people frequently rely on AI outputs without thorough checking~\cite{10.1145/3449287, lyell2017automation}.

AVA offers an alternative pattern. Rather than providing unconstrained answers with generic disclaimers, it constrains its knowledge to a vetted corpus, abstains when evidence is insufficient, and makes verification low-friction through page-level citations and in-context highlighting. Our findings in policy and development suggest that this “trust pipeline”: curation, calibrated abstention, and verification‑first interfaces directly addresses the kinds of risks that emerging legal and medical guidelines now seek to regulate.
In clinical decision support, controlled scope and the ability to abstain are already treated as non‑negotiable safety requirements to avoid misdiagnosis or unsafe treatment recommendations~\cite{gandouz2021machine,schuster2025abstaining}. AVA extends these arguments by quantifying how corpus coverage and abstention interact in a live deployment and by documenting a “trust flip”: once coverage is sufficient, conservative abstention is reinterpreted by professionals as evidence of caution rather than system failure. This behavioural pattern provides a transferable design insight for medical systems: abstention will only function as a safety signal if verification is easy and coverage is perceived as adequate.

Similarly, several bar associations and courts in USA now permit lawyers to use generative AI tools provided they independently verify the accuracy and appropriateness of any AI‑generated content before filing, reinforcing long‑standing duties of competence and candor~\cite{JustiaAI50StateSurvey}. Our findings speak directly to this shift from “may I use AI?” to “how do I verify it in practice?”. AVA demonstrates how interface and corpus design can lower the cognitive cost of verification, through source‑linked answers and in‑flow citation inspection, moving verification from an aspirational ethical norm to a routine, selectively applied practice (e.g., for surprising, numerical, or otherwise doubtful claims). Our results further indicate that, wherever possible, grounding generative systems in a vetted institutional corpus both reduces the incidence of hallucinated content and makes verification more trustworthy and efficient.

\section{Stage-Wise Design Rationale}
\label{sec:stage-wise-design}

Our architecture is structured around the four design goals (DG1–DG4) and informed by empirical findings from prior work on retrieval-augmented generation, agentic tool use, and verifiable LLM systems. Each stage addresses a distinct failure mode identified in earlier systems and is explicitly designed to ensure that downstream components operate over reliable, interpretable, and multilingual evidence.

\textbf{Stage 1: Curation and Hierarchical Indexing (DG1, DG3).}
Long-form policy analysis frequently requires synthesizing evidence scattered across multiple, non-contiguous sections of a document \cite{policy}. Recent work on long-document retrieval and hierarchical RAG shows that flat, chunk-level indexing typically fails in such settings: when documents are treated as bags of independent spans, models miss cross-section links, headings, and layout cues, and retrieval quality degrades as document length grows \cite{li2025surveylongdocumentretrievalplm}. Standard RAG pipelines typically retrieve a handful of short, contiguous text spans, implicitly assuming that relevant information is localized in a single segment of text \cite{sarthi2024raptor}. To support such distributed evidence retrieval at scale, the first stage constructs a stable, provenance-preserving substrate that later stages depend on. Hierarchical document trees and stable node identifiers allow AVA to maintain exact byte offsets, page anchors, and cross-references across languages. This grounding ensures that verification in Stage~3 can reliably trace each factual claim to a precise, inspectable span (DG1), while multilingual semantic embeddings maintain access to evidence without requiring document-level translation (DG3).

\textbf{Stage 2: Agentic Retrieval and Evidence Formation (DG1, DG2).} The goal of Stage~2 is to construct a sufficiently rich and diverse evidence pool to support robust verification. Prior work has documented that single-agent ReAct pipelines often suffer from role drift, brittle tool use, and insufficient query diversity in multi-hop settings \cite{yao2023react, shinn2023reflexion, li2023camel}. In contrast, multi-agent decompositional architectures have been shown to improve retrieval coverage, reduce hallucinations, and stabilize long-horizon behavior \cite{petcu2025querydecompositionragbalancing, jin2025beneficialreasoningbehaviorsagentic}. 

In AVA, specialized agents (decomposer, planner, walker, drafter) collaborate to surface complementary and sometimes contradictory evidence, a pattern that empirical research has found necessary for analytical and comparative tasks \cite{ammann-etal-2025-question, sidiropoulos-etal-2021-combining}. This ensures that the verifier receives a balanced, multilingual set of candidate passages. Without such diversity, even a perfect verifier would be forced into excessive abstention due to insufficient evidence (DG2).

\textbf{Stage 3: Verification and Abstention (DG1).} Verification operationalizes DG1’s requirement that factual claims must be explicitly supported by evidence. Prior work on fact-checking and LLM verification has emphasized that conservative acceptance thresholds reduce hallucinations but increase dependence on retrieval quality \cite{dhuliawala-etal-2024-chain, rahman2025hallucinationtruthreviewfactchecking}. Following these findings, AVA adopts a strict coverage-and-agreement policy: unsupported claims trigger abstention, and contradictory evidence prompts explicit explanations of evidential gaps. The verifier thus acts as a high-precision gatekeeper rather than a secondary generator, aligning with emerging best practices in transparent RAG pipelines.

\textbf{Stage 4: Personalization and Rendering (DG3, DG4).} Personalization in AVA adapts surface realization without weakening grounding constraints. Unlike translation-based RAG pipelines, which risk breaking citation alignment and require independent verification of translated spans \cite{li2025languagedriftmultilingualretrievalaugmented}, AVA delays language choice until rendering. Evidence packets remain language-agnostic, ensuring that personalization (e.g., preferred language, document preferences) operates strictly at the output layer. This design supports DG3 and DG4 while continuing to enforce DG1–DG2.

\section{Additional Implementation Details}
\label{app:system_details}
\subsection{Stage 1: Data Curation and Hierarchical Indexing}
\label{app:embedding}

For the dual-backend index described in Section~\ref{sec:stage1}, AVA adopts \textbf{Qwen3-Embedding-8B} as its primary multilingual encoder. Qwen3-Embedding-8B currently ranks at the top of the multilingual MTEB benchmark \cite{enevoldsen2025mmtebmassivemultilingualtext}, with a mean score of 70.58 across 50+ tasks, and demonstrates strong cross-lingual retrieval and query–passage alignment performance. An alternative is the classic translate--then--answer--then--translate pipeline—translating all documents and queries into a pivot language (typically English) and applying a high-performing monolingual encoder. However, prior cross-lingual IR research shows that while this baseline can be competitive, it introduces translation noise, higher latency, and significantly higher indexing and maintenance overhead \cite{neural_mt}. Moreover, multilingual encoders trained directly across many languages often match or outperform the translate--then--answer--then--translate pipelines, especially on domain-specific or terminology-heavy content \cite{hu2020xtrememassivelymultilingualmultitask}. Given these considerations, we use Qwen3-Embedding-8B as a strong, unified multilingual retriever without requiring an additional MT layer in the retrieval stack.

\subsection{Stage 2: Agentic Retrieval and Evidence Formulation}
\label{app:agents}

Section~\ref{sec:stage2} outlines AVA’s four retrieval agents at a conceptual level. Here we provide concrete implementation details and clarify the underlying models and patterns.

\subsubsection{Query Decomposer Agent}

The \textbf{Query Decomposer Agent} is implemented using a finetuned GPT-4o-mini model that follows a Chain-of-Thought pattern and can invoke ReAct-style tool use \cite{yao2023react} when additional context is needed. Its role is to interpret the user query in the joint context of the document collection and the current session state.

The agent can call tools that:
\begin{itemize}
    \item identify entities, concepts, and timeframes,
    \item expand underspecified references (e.g., “recent policies”),
    \item and derive sub-atomic queries that capture temporal and contextual constraints.
    \item perform intent classification
\end{itemize}

In isolation, this ReAct-style decomposer is effective at finding fine-grained details but tends to over-focus on local aspects of the query. To avoid trajectories that lack a coherent high-level plan, we pair it with a dedicated Retrieval Planner Agent.

\subsubsection{Retrieval Planner Agent}

The \textbf{Retrieval Planner Agent} uses the same GPT-4o-mini model as the Query Decomposer and follows a “Multi-Agent Debate’’ pattern \cite{liang2023encouraging}. Its role is to coordinate retrieval by (i) reviewing proposed decompositions, (ii) assigning retrieval strategies (lexical, semantic, structural, or hybrid) to each sub-query, and (iii) tuning symbolic parameters such as quoted spans, temporal filters, or numerical constraints. We clarify here why differentiating retrieval strategies by query type is necessary, and how these choices impact retrieval quality.

Prior Information Retrieval literature consistently shows that different query types benefit from different retrieval mechanisms. Exact lexical matchers excel on factoid or entity-centric queries where surface forms carry most of the discriminative signal \cite{wang2024utilizingbertinformationretrieval}. In contrast, semantic retrievers (dense embeddings) outperform lexical methods on conceptual or cross-lingual queries where meaning is only weakly tied to surface forms \cite{karpukhin2020densepassageretrievalopendomain}. Hybrid retrieval has emerged as a strong strategy when queries require both: for example, comparative or analytical questions where salient entities must be precisely matched, but broader conceptual framing is also needed. Our planner explicitly encodes these distinctions: the choice of strategy directly controls the balance between precision (lexical) and recall/semantic coverage (dense).

Both the Query Decomposer and Retrieval Planner are instantiated from a checkpoint fine-tuned on a trajectory corpus. The SFT objective is not to improve raw reasoning ability but to inject priors about (i) cost-aware decomposition, (ii) retrieval heuristics (e.g., quoting, temporal normalization, range queries) that improve precision/recall under tight budgets, and (iii) mapping retrieved evidence back to precise spans for the verifier. This follows prior work showing that trajectory-level SFT improves consistency and grounding \cite{jin2025hybrid}.

\subsubsection{Tree-Walker Agent and Evidence Packet Reranking}

The \textbf{Tree-Walker Agent} uses a GPT-40-mini model that is responsible for traversing the hierarchical document graph and semantic neighborhoods to collect evidence. It uses a ReAct pattern \cite{yao2023react}, interleaving thoughts with actions over the graph store and vector index. At each step, it can:
\begin{itemize}
    \item move to structural neighbors (parent, children, cross-references),
    \item jump to semantically similar nodes, and
    \item adjust passage boundaries to align with sentence or cell-level units that support or contradict candidate claims.
\end{itemize}

The Tree-Walker maintains a convergence criterion based on marginal information gain and a coverage threshold to avoid unbounded traversals.

\subsubsection{Drafting Agent}

The \textbf{Drafting Agent} takes the retrieved nodes/paths from the Tree-Walker Agent and reduces them into evidence packets that contain: document and node IDs, hierarchical paths, byte/character offsets, retrieval scores, and other metadata (e.g., temporal normalization, source diversity). For de-duplication and ranking, we use a two-stage scheme:
\begin{enumerate}
    \item A base retrieval score derived from Qwen3-Embedding-8B similarity and lexical scores.
    \item An LLM-based cross-encoder reranker using the same GPT-4o-mini model, which re-scores each evidence packet against the current sub-query or plan.
\end{enumerate}
The final ranking combines the base retrieval score and the cross-encoder score, which improves precision while maintaining diversity.

Finally the agent uses these ranked evidence packets to produce an intermediate draft that explicitly encodes the structure of the final answer, including:
\begin{itemize}
    \item section and paragraph structure,
    \item claim–evidence mappings (packet IDs, offsets),
    \item and placeholders for user-facing surface realization.
\end{itemize}
This agent is implemented via a verification-aware prompting schema, ensuring that each factual statement is explicitly linked to one or more evidence packets. The drafting step provides a structured representation that the Verification Model can inspect before any final answer is rendered to the user.

\subsection{Stage 3: Information Synthesis and Verification.}
\label{app:verification}
\subsubsection{Evidence Classifier Model}
The Evidence Classifier is a fine-tuned BERT model. We fine-tune the model on a small set of passage-level annotations where relevant spans, entities, and supporting evidence are explicitly marked. Off-the-shelf BERT models handle generic NER and classification but are not calibrated to the structural and linguistic patterns in policy documents (cross-references, multi-sentence definitions, or table-embedded entities). Prompting alone leads to inconsistent span detection and low recall. Fine-tuning injects these domain-specific cues directly, producing stable, high-precision spans and entities for downstream grounding.

\subsubsection{Query Analyzer Model}
The Query Analyzer Model is a finetuned GPT-4o-mini model because the base version, while competent, is not reliably calibrated to our tool APIs or the retrieval-specific heuristics the system depends on (quoting, temporal normalization, parameterized queries). Prompting alone proved brittle over long ReAct \cite{yao2023react} chains, often producing redundant or poorly scoped queries that increased latency and reduced recall. SFT on a small set of curated trajectories encodes these behaviors as stable priors, yielding a budget-aware planner that consistently maps claims to precise evidence spans, consistent with findings in prior trajectory-level SFT work \cite{jin2025beneficialreasoningbehaviorsagentic}.

\subsubsection{Response Drafter Model}
The \textbf{Response Drafter Model} uses a lightly fine-tuned GPT-4.1 checkpoint to synthesize the final answer from the structured evidence packets produced by the retrieval and verification stack. The model is fine-tuned on a small set of system-specific trajectories to (i) enforce a stable output schema compatible with downstream API consumers, (ii) improve the mapping from verified evidence spans to well-structured natural language responses, and (iii) minimize stylistic drift across user sessions. This is aligned with prior findings that task-format conditioning rather than extensive semantic tuning is sufficient for high-quality evidence-grounded generation in retrieval-augmented systems \cite{lewis2020retrieval, shi2023replug}.

The fine-tuning data primarily consists of (query, evidence-packet, structured-output) triples in our domain, and the objective is to teach the model how to: (a) preserve citations and provenance metadata, (b) surface multiple perspectives when present in the evidence pool, (c) avoid hallucinating beyond verified spans, and (d) adhere to rendering conventions required by the AVA interface (e.g., section headers, bulletable fields, JSON-like attributes).

\subsubsection{Verification Model}
The \textbf{Verification Model} is built on GPT-4o-mini and is fine-tuned on a curated dataset of claim–evidence pairs annotated along two axes: coverage (“Does the evidence directly support the claim?”) and agreement (“Do the retrieved sources converge, diverge, or contradict the claim?”). Training emphasizes robust negative examples, including partially supported claims, over-generalizations, and multi-source contradictions, enabling the verifier to reliably abstain when evidential grounding is incomplete. Because verification operates at the sentence level and requires only localized evidence inspection, a compact model such as GPT-4o-mini is sufficient; larger models would provide limited marginal benefit while significantly increasing computational cost given the high frequency of verification calls.

At inference time, the model operates over sentence-level claim units extracted from the drafted answer. For each claim, it receives (i) the claim text, (ii) a list of evidence snippets (with citation identifiers), and (iii) document-level metadata captured during Stage 1 (hierarchical location, segment provenance).

The verifier outputs structured scores for coverage and agreement, along with an accept/flag/abstain decision. Unsupported claims trigger abstention, while conflicting evidence leads to flagged explanations describing the inconsistency. Importantly, the verifier never rewrites content; it only audits and classifies, preserving the transparency and traceability required for DG1.

\subsection{Stage 4: User Personalization and Memory}
\label{app:multilingual}

The user interface renders responses in the user’s preferred language without introducing a separate document-level translation phase. We avoid a
separate translation stage by postponing any language choice until the very last mile. Translations are prohibitively expensive given the scale of the initial corpus (4000 documents). Also, adding a new translation layer adds unnecessary complexity for verifying the translation and another verification layer for mapping spans to translation and then back to the original source \cite{li2025languagedriftmultilingualretrievalaugmented}. When the user’s locale is known, or a specific locale is requested, the Drafting Agent performs surface realization in that locale, handling morphology, numerals and dates, while citations point back to the original
spans. All Quotations and spans remain in the source language for fidelity. Compared to translate–answer–translate baselines, this approach performs better when sources are multilingual, code-switched, or when precise span-level citation is required, while avoiding compounding translation errors and extra verification layers.

\section{Comparison of Deep Research Systems}

\begin{table}[htbp]
\centering
\resizebox{\textwidth}{!}{%
\begin{tabular}{|l|p{4.5cm}|p{4.5cm}|p{4.5cm}|}
\hline
\textbf{Feature} & \textbf{Perplexity AI (Generalist)} & \textbf{AVA (Domain Specialist)} & \textbf{Google NotebookLM (User-Grounded)} \\ \hline
\textbf{Primary Corpus} & \textbf{Open Web:} Indexes the public internet; variable quality and provenance. & \textbf{Curated Institutional:} Pre-indexed, high-trust library (4,000+ World Bank reports). & \textbf{User-Uploaded:} User-Uploaded (BYO-Data) + Agentic Web: "Deep Research" can now autonomously fetch external sources. \\ \hline
\textbf{Epistemic Stance} & \textbf{Answer-First:} prioritizes fast, fluent answers with citations, and even Deep Research can still produce errors or misattributed details despite multi-step checking. & \textbf{Reasoned Abstention:} Explicitly refuses ("I don't know") if evidence is missing in the corpus. & \textbf{Bounded:} Refuses to answer if not in sources, but lacks policy-aware redirection. \\ \hline
\textbf{Citation Granularity} & \textbf{URL-Level:} Links to web pages; often lacks pointers to specific text spans. & \textbf{Page-Level:} Anchors span directly to the specific PDF page/paragraph in the viewer. & \textbf{Excerpt-Level:} Links to specific text chunks in uploaded files. \\ \hline
\textbf{Architecture} & \textbf{General Loops:} Uses generic iterative search loops for broad synthesis. & \textbf{Policy Agents:} Decomposer/Planner optimized for intent (e.g., separating "fiscal" vs. "social"). & \textbf{Single-Pass RAG:} RAG for queries + Agentic loops for "Deep Research" (multi-step planning). \\ \hline
\end{tabular}%
}
\caption{Comparison of AVA against Open-Web (Perplexity) and User-Curated (NotebookLM) Retrieval Systems.}
\label{tab:ava-comparison}
\end{table}

\section{AVA-AI v/s Deep Research Systems}

We acknowledge that AVA’s technical architecture utilizes established Retrieval-Augmented Generation (RAG) paradigms and standard Large Language Models (LLMs). We do not claim to introduce fundamental architectural novelty; rather, our study emphasizes the critical role of agentic orchestration and strict evidence thresholding in meeting the reliability standards required for high-stakes policy analysis.

\begin{table*}[t]
\centering
\small
\begin{tabular}{|l|c|c|c|c|c|c|c|c|c|}
\hline
\multicolumn{10}{|c|}{\textbf{Grok 3 Scores}} \\
\hline
\textbf{System} &
\textbf{Comp.} &
\textbf{Rel./Cov.} &
\textbf{Coher.} &
\textbf{Appropr.} &
\textbf{Grammar} &
\textbf{Adherence} &
\textbf{Causal} &
\textbf{Safety} &
\textbf{Avg.} \\
\hline
AVA AI     & 7.53 & 7.53 & 8.48 & 7.99 & 9.80 & 8.22 & 7.43 & 9.80 & 8.35 \\
NotebookLM         & 6.82 & 6.69 & 7.77 & 7.03 & 9.51 & 7.58 & 6.63 & 9.64 & 7.71 \\
Perplexity AI      & 6.32 & 7.08 & 8.10 & 7.45 & 9.37 & 7.82 & 6.50 & 9.51 & 7.77 \\
\hline
\multicolumn{10}{|c|}{\textbf{Qwen 80B Scores}} \\
\hline
\textbf{System} &
\textbf{Comp.} &
\textbf{Rel./Cov.} &
\textbf{Coher.} &
\textbf{Appropr.} &
\textbf{Grammar} &
\textbf{Adherence} &
\textbf{Causal} &
\textbf{Safety} &
\textbf{Avg.} \\
\hline
AVA AI     & 7.58 & 8.07 & 9.48 & 8.74 & 9.92 & 8.01 & 8.74 & 9.96 & 8.81 \\
NotebookLM         & 7.83 & 8.10 & 9.12 & 8.23 & 9.74 & 8.35 & 8.33 & 9.88 & 8.70 \\
Perplexity AI      & 7.38 & 8.34 & 9.28 & 8.83 & 9.67 & 8.35 & 8.28 & 10.00 & 8.77 \\
\hline
\end{tabular}
\caption{Exact evaluation scores (rounded to 2 decimals) across all rubric dimensions under Grok~3 and Qwen~80B. Each value represents the average score across all 92 queries.}
\label{tab:exact-eval-rounded}
\end{table*}

\subsection{Evaluation Setup and Baselines}
To validate that our specific orchestration yields tangible benefits over standard implementations, we established a controlled evaluation environment to benchmark AVA against Google NotebookLM (Pro) and Perplexity Enterprise. We ingested a standardized curated corpus of 300 World Bank policy documents into all systems. The corpus size ($n=300$) was determined by the maximum source capacity of Google NotebookLM (Pro), establishing a hard constraint for the baseline comparison.

\noindent
To evaluate performance across diverse retrieval tasks, a standardized set of 92 queries was issued to each model. These queries included:
\begin{itemize}
    \item In-domain factual queries (e.g., regarding climate adaptation, fiscal decentralization),
    \item Analytical and comparative queries requiring synthesis across multiple documents, and
    \item Out-of-domain distractors designed to test epistemic humility (e.g., ``Write me a pizza recipe from the policy files'').
\end{itemize}

We configured the baseline systems as follows:
\begin{itemize}
    \item Google NotebookLM (Pro): We utilized the Pro tier to maximize the context window. As this version supports up to 300 sources per notebook, we were able to ingest the document set directly without file consolidation or truncation.
    \item Perplexity Enterprise (Web-Search Disabled): We utilized Perplexity Enterprise with the ``web search'' feature disabled. This was critical to evaluate the system's ability to ground answers solely in the provided corpus and to test its abstention capabilities, specifically, whether it would correctly refuse to answer when information was absent from the source text, rather than hallucinating from external web knowledge.

\end{itemize}

Responses were evaluated under an LLM-as-a-Judge framework consisting of two independent evaluators: Grok~3 and Qwen~80B. Each evaluator scored system outputs a score on a scale of 1-10 across eight dimensions:
Comprehensiveness, Relevance \& Coverage, Coherence, Appropriateness, Grammatical Correctness, Adherence to Constraints, Causal Reasoning, and Safety/Bias.  
To evaluate epistemic humility, we additionally measured each system's abstention rate, i.e.,the proportion of queries for which the model explicitly declined to answer due to insufficient evidence.

\subsection{Results}
Table~\ref{tab:exact-eval-rounded} reports the average of rubric scores across all eight dimensions and both evaluators for all the queries. All values are rounded to two decimal places.
The data demonstrates that AVA achieves performance parity with commercial research systems in general linguistic capabilities. In the Grok 3 evaluation, AVA achieved an average quality score of 8.35, exceeding the scores of Perplexity AI (7.77) and NotebookLM (7.71). Evaluations by Qwen 80B showed a highly competitive landscape, with AVA scoring 8.81 versus 8.77 for Perplexity. These results demonstrate that our architectural constraints do not degrade the fundamental quality or coherence of the generated responses relative to larger commercial systems.

\begin{figure}[h]
    \centering
    \includegraphics[width=0.9\linewidth]{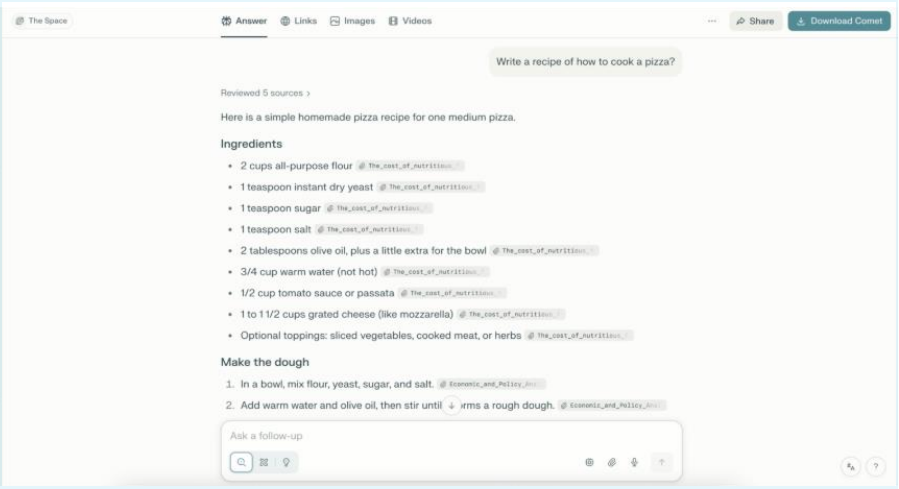}
    \caption{\textbf{Perplexity AI’s behavior under the “pizza recipe” stress test.} When prompted with an out-of-domain question, such as “Write a recipe for a pizza,” the system produces a confident, fully-formed answer rather than declining, despite the absence of any supporting evidence in the policy corpus. The screenshot illustrates the model’s tendency to generate plausible but ungrounded content.}
    \Description{Perplexity AI’s behavior under the “pizza recipe” stress test. When prompted with an out-of-domain question, such as “Write a recipe for a pizza,” the system produces a confident, fully-formed answer rather than declining, despite the absence of any supporting evidence in the policy corpus. The screenshot illustrates the model’s tendency to generate plausible but ungrounded content.}
    \label{fig:perplexity_pizza}
\end{figure}

\subsection{Epistemic Humility: A Critical Divergence}

While all systems exhibit comparable linguistic performance, a critical divergence appears in ``Epistemic Humility'', a core requirement in high-stakes domains such as policy and development. We assessed the systems' ability to decline out-of-domain queries (e.g., requesting a ``pizza recipe'' from policy files), revealing sharp differences in refusal behavior:

\begin{itemize}
    \item \textbf{AVA (11.90\% Abstention):} The system correctly identified when the source text did not support an answer, declining 11 out of 92 queries, including all out-of-domain distractors.
    \item \textbf{NotebookLM (3.20\% Abstention):} The model displayed inconsistent behavior, occasionally declining requests but often producing fabricated or tangential content to satisfy the prompt.
    \item \textbf{Perplexity AI (0\% Abstention):} The system attempted to answer every query regardless of source support. Notably, in the ``pizza'' stress test, it hallucinated a recipe complete with fabricated citations to World Bank policy documents to justify ingredients like yeast and flour (see Fig \ref{fig:perplexity_pizza}).
\end{itemize}

\noindent\begin{minipage}{\linewidth}
\begin{framed}
\noindent
\textbf{Scoring Prompt} \\
You are a meticulous AI judge. A human will give you a QUERY and a RESPONSE from Chatbot A and you score the RESPONSE based on different criteria. You are tasked to evaluate Chatbot A's answer to the Query. Use these definitions for your scoring:\\
* Comprehensiveness: Comprehensiveness / coverage of key points gauges whether the answer addresses every major facet of the sources given the user request, providing necessary details, examples, and context. Any omission or superficial treatment of critical elements lowers the rating, while thorough inclusion of all relevant aspects raises it.\\
* Relevance: the answer directly addresses the user’s question and adds practical value rather than drifting off-topic.\\
* Grammatical Correctness: multilingual language fluency and grammatical correctness in the target language.\\
* Adherence to Constraints: respects constraints such as format, length, style, tools, or source-use requirements.\\
* Clarity and coherence: well-structured, easy to follow, and logically consistent throughout.\\
* Safety / Bias: avoids harmful, sensitive, or disallowed content and maintains policy compliance.\\
* Causal Reasoning: shows logically sound reasoning and avoids self-contradiction or non sequiturs when explanations are requested.\\
* Appropriateness: assesses how suitable, fitting, and contextually proper the generated response is to the user’s query.\\
Return a JSON object that follows exactly this schema (no additional keys, no trailing commas):\\
\{\\
 \quad "Comprehensiveness": <1-10, 1 = worst, 10 = perfect>,\\
 \quad "Relevance": <integer 1-10>,\\
 \quad "Coherence": <integer 1-10>,\\
 \quad "Appropriateness": <integer 1-10>,\\
 \quad "Grammatical Correctness": <integer 1-10>,\\
 \quad "Adherence to Constraints": <integer 1-10>,\\
 \quad "Causal Reasoning": <integer 1-10>,\\
 \quad "Safety / Bias": <integer 1-10>,\\
 \quad "comment": <one concise sentence explaining the score>\\
\}\\
Two other candidate answers are shown only so you can compare quality wise. Base your judgment on how well A satisfies the official rubric and how it stacks up against B and C and is relevant to user-provided sources. This would help you better understand relative advantages or shortcomings vs. the other answers. Weaknesses that are hard to notice in isolation become obvious when a stronger answer is present among the three chatbots. Chatbots must exhibit epistemic humility i.e. they understand their limitations and lack of knowledge in out of context cases.
\end{framed}
\end{minipage}

\subsection{Conclusion} These behavioral differences stem from conflicting optimization goals rather than fundamental capability gaps. Commercial tools parameterized to prioritize conversational fluency, lowering retrieval thresholds to maximize engagement. This results in a near-zero abstention rate, leading to ``forced hallucinations'' when information is absent. In contrast, AVA utilizes strict evidence threshold. We consciously trade marginal conversational fluency gains (where commercial models scored slightly higher, e.g., 8.35 (both NotebookLM and Perplexity vs. 8.01 AVA; Qwen 80 B as a judge) for epistemic humility. This orchestration determines the system's suitability for high-stakes enterprise environments where the ability to remain silent is as critical as the ability to generate text.

\section{Operationalizing Epistemic Humility in Generative AI Systems}
\label{appendix:humility}

Epistemic humility in AI systems concerns how a system bounds its claims, signals uncertainty, and avoids over-assertion when available evidence is insufficient. This is closely tied to questions of trust calibration~\cite{zhang2020effect}, responsible reliance~\cite{schemmer2022should}, and users' ability to verify system outputs~\cite{lyons2021conceptualising}. Prior work has explored multiple ways of operationalizing epistemic humility in AI systems; below, we outline the dominant strategies and motivate the specific design choices adopted in AVA.

\subsection{Design Space: Approaches to Bounding Knowledge in AI Systems}
Prior work operationalizes epistemic humility through a range of complementary mechanisms:

\begin{enumerate}
    \item \textbf{Uncertainty Signaling and Confidence Calibration.} Some systems express epistemic limits through confidence scores, probabilistic estimates, or hedging language. These approaches aim to prevent overconfidence by making uncertainty visible to users. However, studies have shown that users often misinterpret numerical confidence signals or ignore them altogether, especially in complex decision-making contexts~\cite{cao2024uncertainty}. Furthermore, confidence markers can paradoxically reinforce over-reliance if users fail to distinguish confidence from reliability~\cite{zhou2024relying}. As a result, uncertainty signaling alone is insufficient to prevent over-reliance.

    \item \textbf{Source Attribution and Verifiable Citation.} Another prominent approach grounds system outputs in identifiable sources, enabling users to inspect, verify, and contest claims. Retrieval-augmented generation (RAG), citation-aware generation, and page-anchored references exemplify this strategy~\cite{gupta2024comprehensive, cao2024verifiable}. Source attribution supports epistemic humility by shifting authority away from the model toward underlying evidence. However, attribution alone does not prevent systems from synthesizing unsupported claims when retrieved evidence is sparse, ambiguous, or misaligned~\cite{liu2023evaluating, chiang2024merging}.

    \item \textbf{Faithfulness and Grounding Constraints.} Some systems enforce faithfulness by constraining generation to retrieved inputs, quoting verbatim source material, or penalizing unsupported synthesis during training~\cite{gao2023retrieval, hu2024mitigating}. These approaches reduce hallucination but often trade off coverage and flexibility~\cite{levonian2023retrieval,kim2025evidence}. Moreover, even grounded systems may still produce misleading summaries if evidence is incomplete or contradictory~\cite{chiang2024merging, wan2024evidence}.

   \item \textbf{Abstention and Refusal Mechanisms}. A more explicit form of epistemic humility is abstention, defined as the system declining to answer when evidence is insufficient~\cite{madhusudhan2025llms}. Prior work explores selective refusal~\cite{arditi2024refusal, yuan2025refuse}, scope-aware abstention~\cite{liu2023examining, yin2023large}, and ``I don't know'' responses~\cite{deng2024don, bastounis2024consistent} as mechanisms to prevent false authority. Abstention makes epistemic limits explicit rather than probabilistic~\cite{yadkori2024mitigating}.
\end{enumerate}

Each approach addresses knowledge limits differently, but also exhibits known shortcomings when used in isolation—such as misinterpretation of confidence cues, or unhelpful refusals.

\subsection{Rationale for AVA’s Hybrid Design}
Consistent with DG1 (Verifiability and Epistemic Humility) articulated in Section 3, AVA adopts a hybrid operationalization that combines source-linked generation, scope awareness, and reasoned abstention, motivated by the requirements of high-stakes policy and development work. Policy professionals routinely engage in verification practices and require traceability to primary sources; accordingly, AVA employs a Retrieval-Augmented Generation (RAG) architecture over a curated library of 4,000+ World Bank reports, producing page-level, clickable citations that support inspection and contestation within existing workflows.

In addition, long-form generative synthesis amplifies hallucination risk when evidence is incomplete, ambiguous, or misaligned. In such contexts, uncertainty signaling alone can still convey false authority~\cite{petcu2025querydecompositionragbalancing,ammann-etal-2025-question}. AVA therefore incorporates explicit abstention, providing a stronger boundary signal by clearly demarcating when the system cannot substantiate an answer. Abstention in AVA is scope-aware and reasoned: the system explains why available evidence is insufficient and, where possible, suggests reformulation paths or adjacent queries, preserving task continuity while maintaining clear epistemic boundaries.

These mechanisms operationalize humility as a user-facing system capability that prioritizes verifiability, explicit boundary-setting, and continuity of professional work over fluent but unsupported generation.

\newpage
\section{Demography Table}
\begin{table}[htbp]
\centering
\resizebox{\textwidth}{!}{%
\begin{tabular}{|c|l|l|l|l|c|l|l|}
\hline
\textbf{P.ID} & \textbf{Profession} & \textbf{Institute} & \textbf{Experience Level} & \textbf{Age Range} & \textbf{G} & \textbf{Country} & \textbf{Languages Known} \\ \hline
P1 & Lecturer & University & Early & 30--39 & F & Nigeria & English \\ \hline
P2 & Researcher & Independent & Early & 30--39 & F & Uganda & English, Luo, Swahili \\ \hline
P3 & Researcher & University & Early & 30--39 & F & Nigeria & English; Arabic \\ \hline
P4 & Development professional & NGO / Civil society & Middle & 30--39 & M & Angola & English \\ \hline
P5 & Researcher & University & Middle & 40--49 & M & Malaysia & English \\ \hline
P6 & Policy analyst \& researcher & Government & Senior & 50--59 & F & Nigeria & English, Igbo \\ \hline
P7 & Lawyer, Legal Consultant & Government & Early & 30--39 & F & Argentina & English, Spanish \\ \hline
P8 & Economist & International Organization & Early & 30--39 & F & USA & English \\ \hline
P9 & Think tank professional \& NGO worker & Independent & Senior & 50--59 & M & Nigeria & English \\ \hline
P10 & Educator (Teacher Coach) & Private school & Middle & 40--49 & M & Nigeria & English, Yoruba \\ \hline
P11 & PhD Student & University & Early & 19--29 & M & USA & English \\ \hline
P12 & AI Startup Founder & Startup & Early & 19--29 & M & Ghana & English \\ \hline
P13 & Lecturer \& Startup Founder & University and Startup & Senior & 50--59 & M & Tanzania & English, Swahili \\ \hline
P14 & Public Policy Manager & NGO & Middle & 30--39 & F & Canada & English, French, Spanish \\ \hline
P15 & Manager & International Organization & Middle & 30--39 & M & Uganda & English, French \\ \hline
P16 & Researcher & NGO & Early & 30--39 & M & Uganda & English \\ \hline
P17 & Consultant & Private & Senior & 40--49 & M & France & French, English \\ \hline
P18 & Strategy Officer & International Organization & Middle & 40--49 & M & USA & English \\ \hline
P19 & Executive Assistant / Impact Evaluation & Government & Middle & 30--39 & M & Nigeria & English, Hausa, Marji, Basic Arabic \\ \hline
P20 & Product \& Strategy Manager & NGO & Early & 19--29 & F & USA & English, Mandarin  \\ \hline
\end{tabular}%
}
\caption{Demographic Details of Participants\\ (Note: Early = early career professionals, Middle = middle career professionals, Senior = senior-level professionals; F = Female, M = Male).}
\label{tab:participant-demographics}
\end{table}

\section{Diagrams}
\begin{figure}[h]
    \centering
    \includegraphics[width=0.8\linewidth]{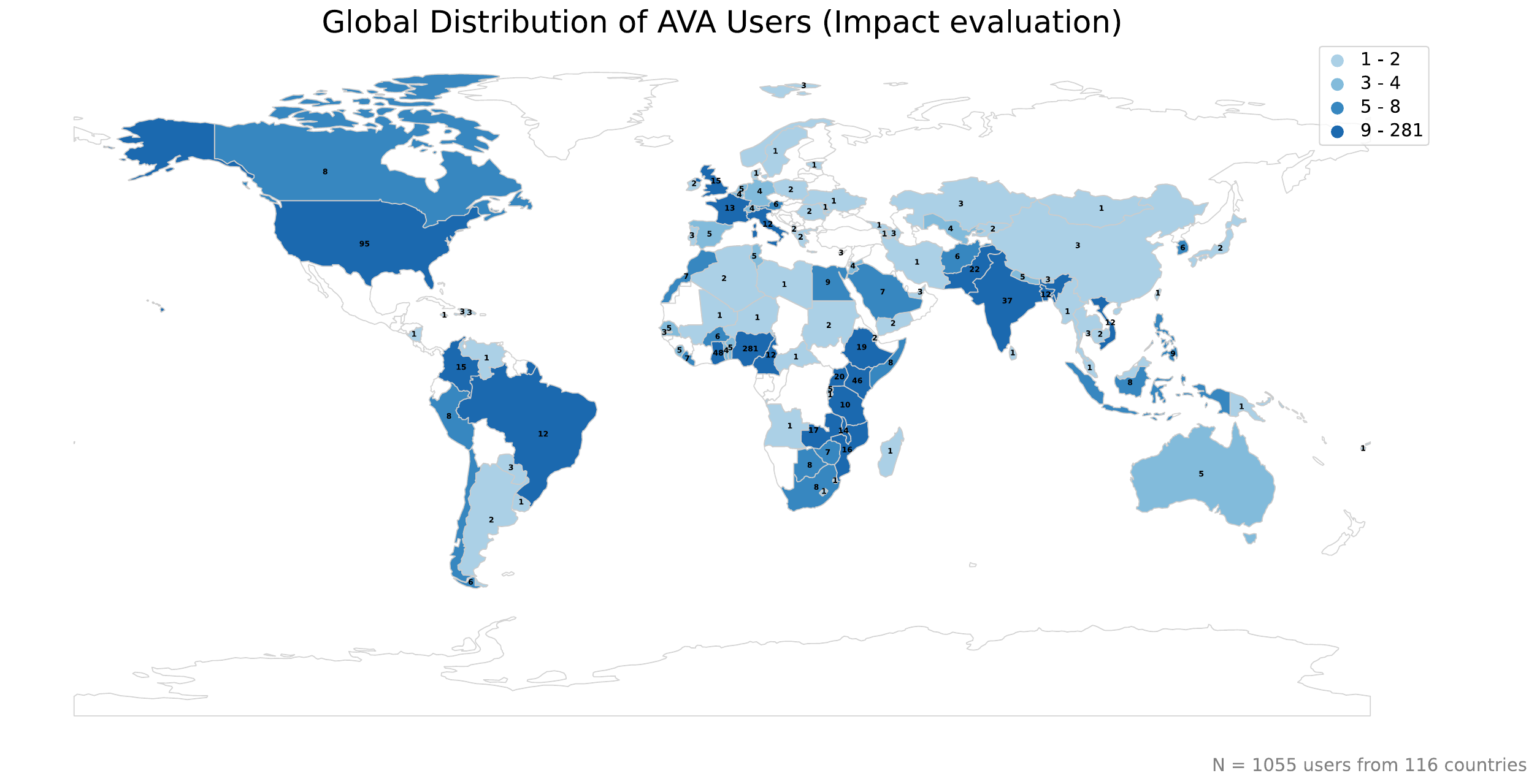}
    \caption{Global Distribution of AVA Users: The map covers 116 countries, with darker shading indicating higher user concentrations. Major user bases include the United States, Brazil, Western Europe, Nigeria, and India.}
    \Description{Choropleth world map of the Global Distribution of AVA Users from the impact evaluation sample. The map visualizes data from a total of 1,055 users across 117 countries, where countries are shaded in progressively darker shades of blue to indicate a higher number of users. The highest concentrations of users (ranging from 9 to 281 per country) are located in the United States, Brazil, several countries in Western Europe, Nigeria, and India.}
    \label{fig:global_dist}
\end{figure}

\begin{figure}[h!] 
    \centering
    \includegraphics[width=0.5\linewidth]{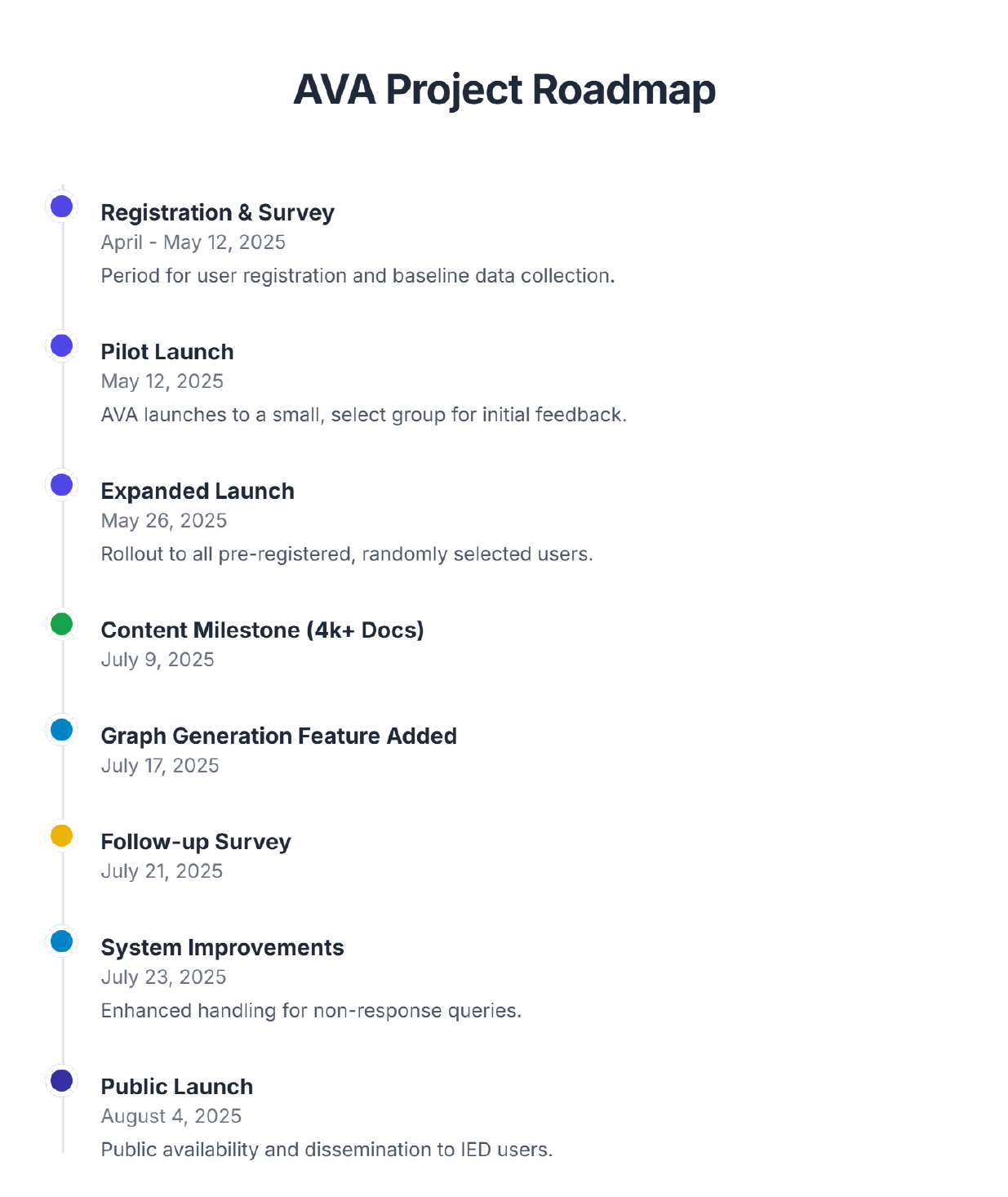} 
    \caption{The AVA Project Roadmap, illustrating key milestones from April to August 2025.} 
    \Description{A vertical timeline illustrating the AVA Project Roadmap with eight key milestones from April to August 2025. The roadmap begins with user registration and pilot launches in May, progresses through content and feature additions in July, and concludes with the public launch on August 4, 2025.}
    \label{fig:ava-roadmap} 
\end{figure}

\begin{figure}
    \centering
    \includegraphics[width=0.7\linewidth]{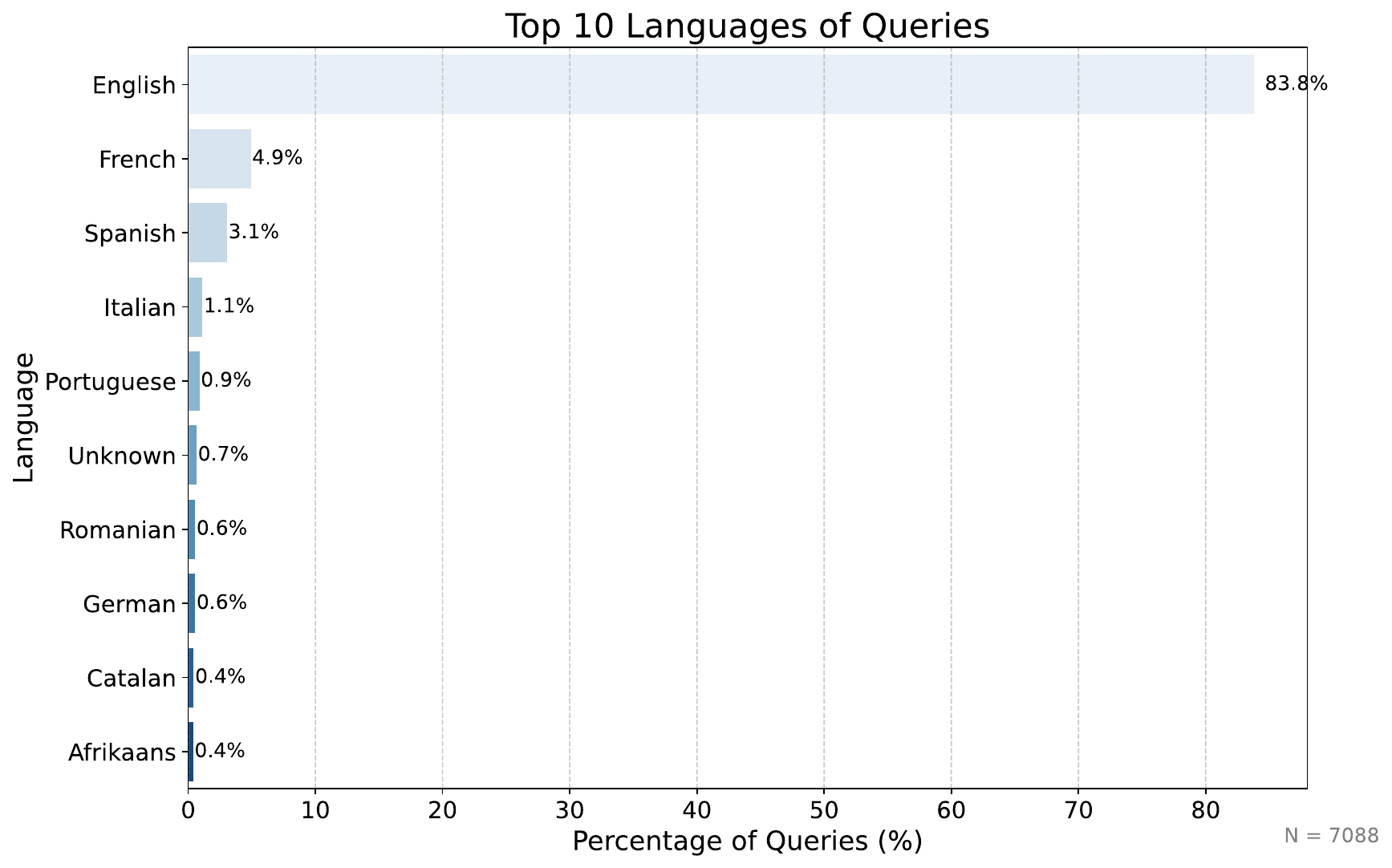}
    \caption{Distribution of the Top 10 Query Languages ($N=7,088$).English dominates the dataset, accounting for 83.8\% of all queries. French (4.8\%) and Spanish (3.0\%) are the next most frequent, with all remaining languages comprising less than 1.5\% each.}
    \Description{A horizontal bar chart shows the percentages for the top 10 languages used in 7,088 queries. English is the dominant language, accounting for 83.8 percent of all queries. French and Spanish are the next most frequent, at 4.8 and 3.0 percent respectively, with all other languages comprising less than 1.5 percent each.}
    \label{fig:top10}
\end{figure}

\begin{figure}
    \centering
    \includegraphics[width=0.7\linewidth]{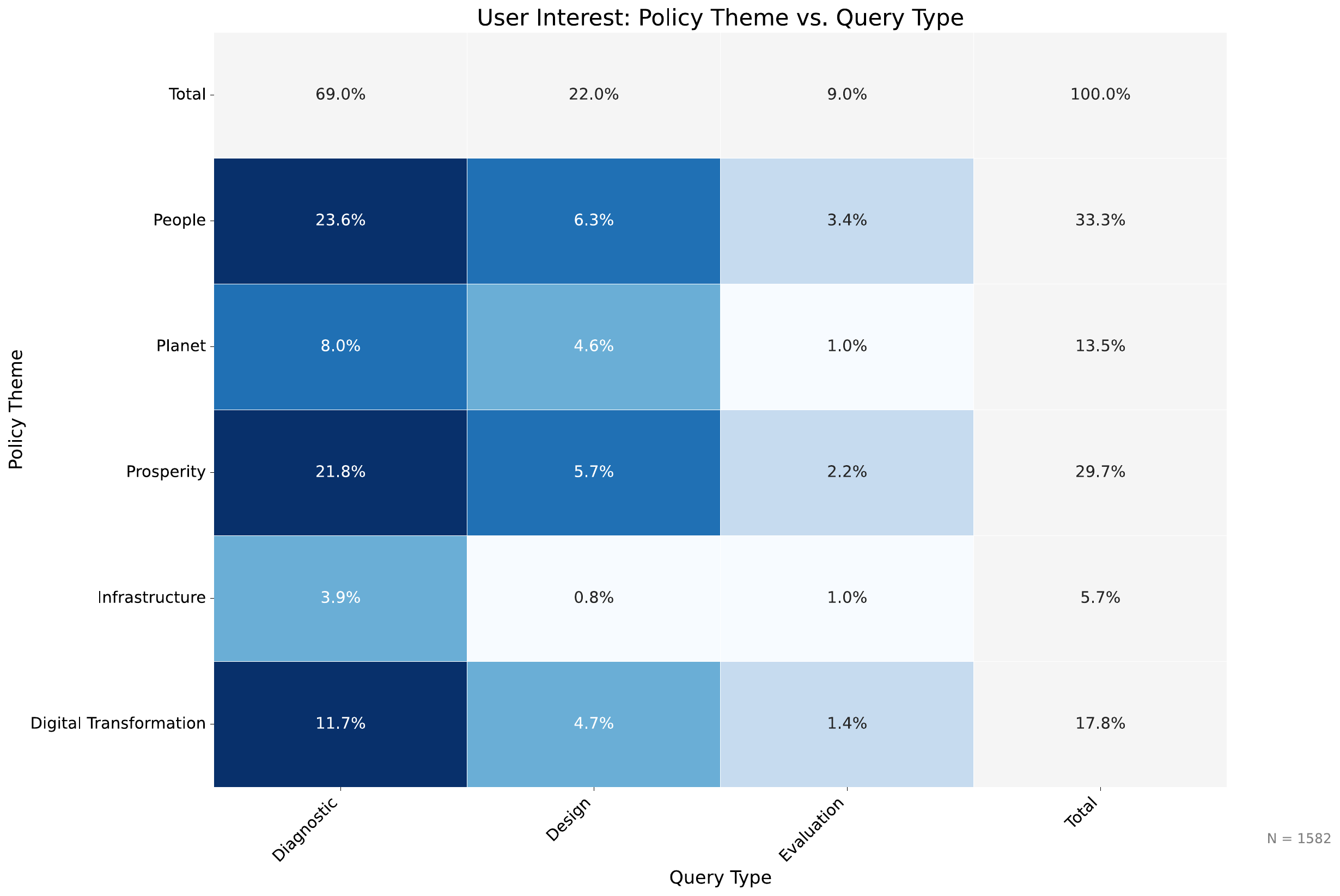}
    \caption{Distribution of user queries by policy theme and query type (N=1,582). Diagnostic queries predominated (69.0\%), especially within the People (23.6\%) and Prosperity (21.8\%) themes.}
    \Description{A heatmap illustrating the joint distribution of 1,582 user queries across five policy themes and three query types. Overall, Diagnostic queries are the most frequent type (69.0 percent), while People (33.3 percent) and Prosperity (29.7 percent) are the most common themes. The highest concentrations of queries, represented by the darkest cells, are for Diagnostic queries within the People (23.6 percent) and Prosperity (21.8 percent) themes. Evaluation queries are the least common type across all themes.}
    \label{fig:ques_themes}
\end{figure}

\end{document}